\documentclass[12pt]{article}
\usepackage[utf8]{inputenc}
\usepackage[T1]{fontenc}
\usepackage{graphicx}
\usepackage{placeins} 
\usepackage{amsmath}
\usepackage{authblk}
\usepackage{caption}

\usepackage{geometry}
\usepackage{setspace}
\begin{document}

\title{High-throughput viscometry via machine-learning from~videos~of inverted vials}

\author{Ignacio Arretche$^{\text{a,1,2}}$, Mohammad Tanver Hossain$^{\text{a,b,1}}$, Ramdas Tiwari$^{\text{a,b}}$, Abbie Kim$^{\text{a,b}}$, Mya G. Mills$^{\text{c}}$, Connor D. Armstrong$^{\text{a,d}}$, Jacob J. Lessard$^{\text{e}}$, Sameh H. Tawfick$^{\text{a,b,2}}$, and Randy H. Ewoldt$^{\text{a,b,2}}$ }
\date{}  
\maketitle

\noindent$^{\text{a}}$Beckman Institute for Advanced Science and Technology, UIUC, Illinois 61801\\
$^{\text{b}}$Department of Mechanical Science and Engineering, UIUC, Illinois 61801\\
$^{\text{c}}$Department of Chemistry, UIUC, Urbana, Illinois 61801\\
$^{\text{d}}$Department of Material Science and Engineering, UIUC, Urbana, Illinois 61801\\
$^{\text{e}}$Department of Chemistry, University of Utah, Salt Lake City, Utah 84112\\
$^{\text{1}}$ IA and MTH contributed equally\\
$^{\text{2}}$To whom correspondence should be addressed.\\ E-mail: ia6@illinois.edu, tawfick@illinois.edu and ewoldt@illinois.edu





\footnotesize
\begin{abstract}
Although the inverted vial test has been widely used as a qualitative method for estimating fluid viscosity, quantitative rheological characterization has remained limited due to its complex, uncontrolled flow - driven by gravity, surface tension, inertia, and initial conditions. Here, we present a computer vision (CV) viscometer that automates the inverted vial test and enables quantitative viscosity inference across nearly five orders of magnitude (0.01–1000~Pa$\cdot$s), without requiring direct velocity field measurements. The system simultaneously inverts multiple vials and records videos of the evolving fluid, which are fed into a neural network that approximates the inverse function from visual features and known fluid density. Despite the complex, multi-regime flow within the vial, our approach achieves relative errors below 25\%, improving to 15\% for viscosities above 0.1~Pa$\cdot$s. When tested on non-Newtonian polymer solutions, the method reliably estimates zero-shear viscosity as long as viscoelastic or shear-thinning behaviors remain negligible within the flow regime. Moreover, high standard deviations in the inferred values may serve as a proxy for identifying fluids with strong non-Newtonian behavior. The CV viscometer requires only one camera and one motor, is contactless and low-cost, and can be easily integrated into high-throughput experimental automated and manual workflows. Transcending traditional characterization paradigms, our method leverages uncontrolled flows and visual features to achieve simplicity and scalability, enabling high-throughput viscosity inference that can meet the growing demand of data-driven material models while remaining accessible to lower resource environments.

\end{abstract}

\normalsize

\section*{Introduction}
The field of material science is experiencing a paradigm shift, driven by the use of data-driven techniques that can predict material properties and uncover unprecedented trends. In contrast to the historically linear process of sequential testing and analysis, which relied heavily on scientific intuition, modern approaches leverage machine learning algorithms to mine vast datasets and directly uncover multivariable nonlinear correlations~\cite{1,2,3,4}. These data-driven techniques are unlocking entirely new ways for functional material discovery at a scale and speed previously thought impossible. However, the success of these methods hinges on the availability of large data sets. Massive amounts of high-quality, diverse data are required to fuel these data-driven models, considerably increasing the need for high-throughput material characterization.

\begin{figure*}
\centering
\includegraphics[width=17.8cm]{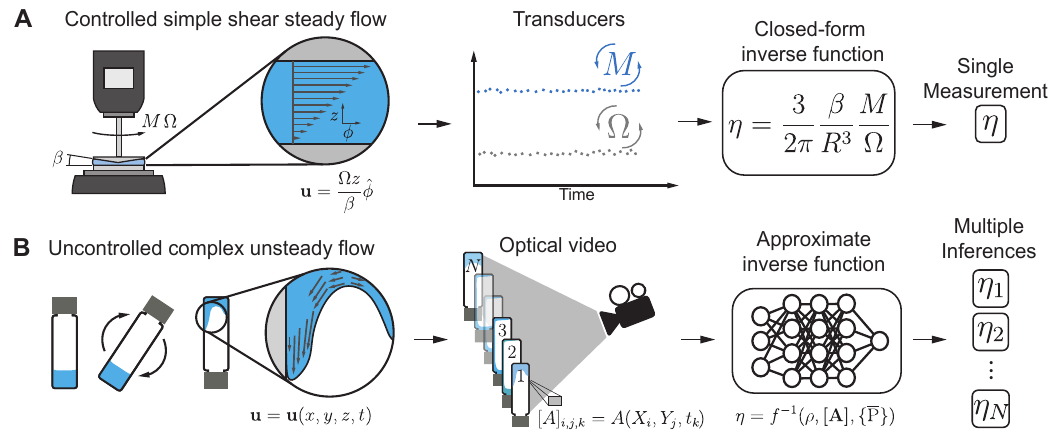}
\caption{Workflows for (A) Measurement of viscosity using small angle cone and plate simple shear steady flow. (B) Inference of viscosity from the complex flow of an inverted vial test.}
\label{fig:fig1}
\end{figure*}

Traditional characterization techniques, while accurate, are typically incompatible with the demands of modern material discovery. They are often too slow to generate the required volume of data, or they rely on expensive, labor-intensive equipment to maintain the controlled environments typical of these tests. This can be particularly evident in the measurement of rheological properties, which are essential for developing suitable material processing conditions~\cite{5}, designing a wide range of consumer products~\cite{6}, and unraveling the chemical kinetics of reactive systems~\cite{7}. Conventional shear rheometry, e.g.~using parallel plate or cone and plate (Fig.~\ref{fig:fig1}A), employs controlled velocity fields to achieve precise measurements through closed-form inverse solutions and high-accuracy transducers. However, maintaining these specific fields imposes considerable operational challenges~\cite{8}, including precision machine design, complex sample preparation, the use of disposable components, rigorous cleaning cycles, and the need for delicate transducers. While automated rheometers based on these controlled flows exist, they are still slow, limited by the sample preparation steps and the typically long measurement cycles, and rely on expensive robotic interfaces~\cite{9} or costly micro-sensing technologies~\cite{10,11}. Automation enhances their usability and efficiency in daily operations; however, their throughput, defined as tests per unit time, remains limited. Achieving materials characterization at a rate of over $10^3$ tests per day requires new disruptive testing methodologies, as traditional techniques were designed with accuracy as the primary objective rather than throughput.

In response to these challenges, the use of complex but visually accessible flow fields has emerged as a promising alternative for rheological characterization. A broad collection of these tests has recently been introduced under the concept of ``protorheology''~\cite{12}. Protorheology~\cite{12} enables the inference of rheological properties from imprecise, uncontrolled, or non-ideal deformations. The typically less-controlled environment of these tests makes them cost-effective, accessible, fast, and simpler to automate~\cite{12,13,14,15,16,17,18}. These methods do not match the accuracy of a precision rheometer but offer immense gains in efficiency and scalability. Yet, the complexity of the flow field comes with associated challenges when it comes to quantitative inference of rheological properties. 

The tilted or inverted vial test is one of the most widely used protorheology tests for qualitative rheological understanding~\cite{7,19,20,21,22,23,24,25,26}, however, its complex flow physics has inhibited quantitative rheological inference. In this test, a vial containing the fluid of interest is either tilted or fully inverted (180-degree tilt) to observe the fluid flow inside the vial. For tilt angles less than 90 degrees, gravity causes the surface to level and the viscosity can potentially be related to a characteristic leveling time~\cite{12,17}. However, the flow behavior in the fully inverted vial is far more complex, which makes this approach unlikely to yield accurate viscosity estimates. 

Despite its widespread use, there is no analytical or numerical approximation to the full-flow development of the forward (nor inverse) problem for the fully inverted vial test. The fluid undergoes multiple 3D flow regimes as the test progresses, covering a wide range of Reynolds (Re) and Bond (Bo) number and making it impossible to develop an analytical solution that can capture all regimes. The inverted vial test remains neither fully automated nor quantitatively understood.

In complex flows like this one, machine learning offers a compelling alternative for approximating the inverse problem. Data-driven approaches have been applied effectively to infer complex rheological parameters~\cite{28,29,30,31} from traditional flows such as simple shear (Fig.~\ref{fig:fig1}A). Machine learning can also be used to solve inverse problems of complex flows. Promising results have emerged from neural networks trained on simulated data of complex flows, such as pouring and stirring~\cite{32}, vortex streets~\cite{33}, or microfluidic particle trains~\cite{34}. Experimental non-Newtonian rheological inference has also been achieved using flow-MRI measurements~\cite{35}, video-based analysis of pendant drops~\cite{36} and inclined plane flows~\cite{37}, and image-based analysis of direct ink writing patterns~\cite{38}. These studies showcase the potential of rheological inference from spatiotemporal data, but some of the flows they analyze may be difficult to automate in a high-throughput laboratory environment, and they only estimate viscosities (or model parameters) in ranges of about one to two orders of magnitude.  A recent study employed a 3D-CNN to classify fluids into five viscosity ranges (0.3~Pa$\cdot$s to 20~Pa$\cdot$s) based on videos of vials tilted 90 degrees by a robotic arm~\cite{16}. While this work successfully demonstrated fluid classification from the videos, the range of viscosities was also limited to two orders of magnitude and it only offers limited insights into regression models that can quantify viscosity values using this flow. Moreover, the use of a robotic arm restricts accessibility and scalability for parallel testing.

Here, we focus on the fully inverted vial test. Its simplicity makes it accessible and allows for simultaneous multiplexed inference. We hypothesize that full inversion allows for the inference of a broad range of viscosities due to the development of multiple flow regimes. The velocity field of the liquid in an inverted vial test can be expressed as
\begin{equation}
\textbf{u}(x,y,z,t)=g(\{\text{P}\})
\end{equation}
where the function $g()$ depends on \{P\}, the collection of all information necessary to define the flow including but not limited to constitutive behavior, fluid density, surface tension, contact angles, domain geometry, boundary and initial conditions, and forcing - the latter of which are defined by vial size, fluid volume, and rotating speed. Assuming a Newtonian fluid with constant viscosity $\eta$ and density $\rho$,
\begin{equation}
\textbf{u}(x,y,z,t)=g(\eta, \rho, \{\overline{\text{P}}\})
\label{eq:forward}
\end{equation}
where $\{\overline{\text{P}}\}$ is \{P\} excluding the constitutive behavior and density. If the measured flow is sensitive to viscosity, e.g.\ viscous flow, then we can invert Eq.~\ref{eq:forward} to find
\begin{equation}
\eta = g^{-1}(\textbf{u}(x,y,x,t), \rho, \{\overline{\text{P}}\}).
\end{equation}

We could calculate $g^{-1}$ using Bayesian approaches~\cite{39,40} or neural networks~\cite{33}, among other methods. However, inferring $\eta$ from $g^{-1}$ requires accurate \(\textbf{u}\) measurements, typically made through complex techniques like flow-MRI~\cite{40}, particle tracking, or Doppler ultrasonic velocimetry~\cite{41}. 

Instead, we propose observing the flow field from video recordings, simply capturing the optical features which are seen from the front view of the vials, $[A]_{i,j,k}(X_i, Y_j, t_k)$ (Fig.~\ref{fig:fig1}B), which include optical distortions due to curvature of the vial and free surface of the fluid. This measurement is only a proxy of the flow field. For example, the most apparent feature may be the displacement of the projected meniscus. Hence, \textbf{[A]} is not a full-field flow tensor and does not require complex measurement and computation methods. Rather, it is captured using a video camera, making the method highly accessible and suitable for a neural network framework.

We hypothesize that if the transient flow dynamics in an inverted vial test are sufficiently sensitive to viscosity, we can infer the viscosity of a Newtonian fluid with known density by using video recordings of the transient flow of an inverted vial test with testing parameters, $\overline{\text{P}}$, and a supervised neural network trained with videos of tests under those same testing parameters. That is,
\begin{equation}
\eta = f^{-1}(\rho, \textbf{[A]}, \{\overline{\text{P}}\})
\label{inverse fucntion}
\end{equation}
where \textbf{[A]} is a matrix that represents optical features captured from the video recordings, and ${f}^{-1}$ is an inverse function approximated by the neural network. The set $\{\overline{\text{P}}\}$ includes operational parameters, such as the vial size and fluid volume, and operational functions, such as the tilting angular speed profile used in the test. These operational inputs are controlled and kept constant for training and testing. The $\{\overline{\text{P}}\}$ also includes fluid properties such as surface tension and contact angle, which we do not measure nor control.  For accurate viscosity inference, these fluid-dependent properties must either remain similar across training and testing fluids or have a negligible influence on the transient flow dynamics.

\begin{figure}
\centering
\includegraphics[width=8.9cm]{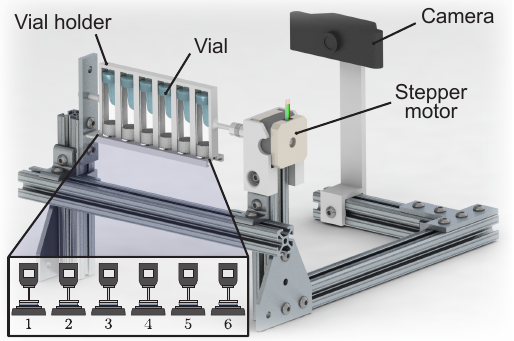}
\caption{Schematic of the computer vision (CV) viscometer for 6 parallel measurements of viscosity.}
\label{fig:protoviscometer}
\end{figure}

This hypothesis is non-trivial because (1) it relies on the existence of an inverse function and (2) video recordings do not provide full velocity field measurements but instead capture optically contrasted flow features, influenced by factors such as lighting, refraction, and the source of optical contrast. However, if successful, a single video of multiple inverted vials could enable multiplexed inference, allowing for parallelization and increased throughput (Fig.~\ref{fig:fig1}B). 

By leveraging simple, high-throughput tests with multiplexing capabilities, we propose a paradigm shift in material characterization. Rather than relying on closed-form inverse solutions for controlled flows with complex procedures (Fig.~\ref{fig:fig1}A), we harness machine learning to infer viscosity from complex, video-captured flow patterns, enabling fast, simple, and high throughput experimental characterization (Fig.~\ref{fig:fig1}B).

\section*{Results}
\subsection*{Experimental setup: the computer vision (CV) viscometer}

To acquire the necessary data, we designed an automated system, which we refer to as a computer vision (CV) viscometer. The CV viscometer inverts an array of vials (Fig.~\ref{fig:protoviscometer}) following a programmed sequence and records videos of the fluid flow. The procedure starts by filling vials with the liquids of interest, then the vials are placed upright in a rack. The program sends the command to flip the rack $180^{\circ}$  by a stepper motor within a duration $t_{\text{flip}}$. A video of the flipped vial is recorded for a fixed observation time, $t_{\text{obs}}$. The vials are flipped back to the upright position after $t_{\text{obs}}$. The only sensor of the CV viscometer is a camera. 

\subsection*{The neural network structure}
\begin{figure*}
\centering
\includegraphics[width=17.8cm]{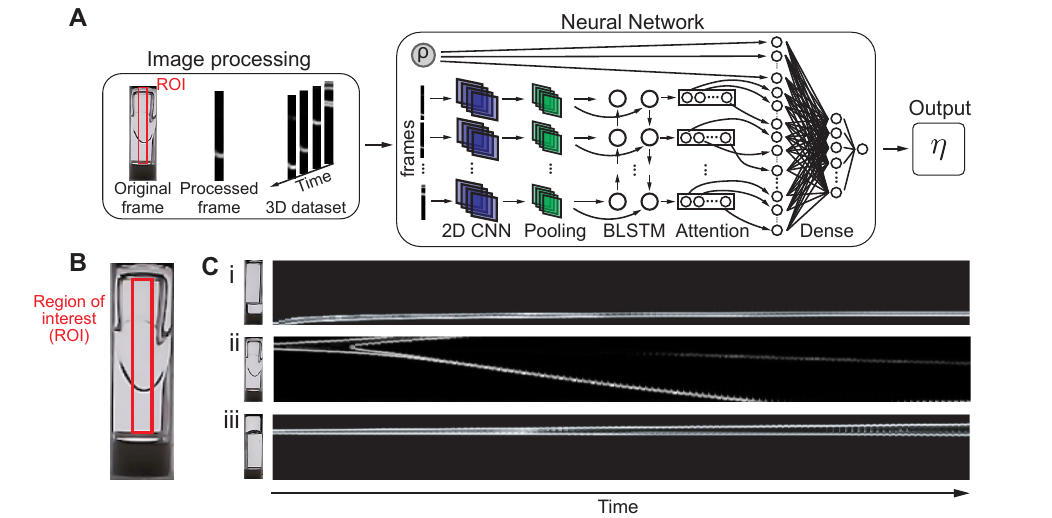}
\caption{Videos and image processing of the inverted vial test. (A) Schematic of the image processing and neural network structures. (B) Region of interest. (C) Image of fluid 30 seconds after flip (left) and the 112 concatenated frames after image processing (right) for fluids with shear viscosities equal to 0.7~Pa$\cdot$s (i), 30~Pa$\cdot$s (ii), 137~Pa$\cdot$s (iii).}
\label{fig:NN}
\end{figure*}

Our neural network architecture is inspired by the promising results of supervised Convolutional Neural Networks (CNNs) and Recurrent Neural Network (RNN) structures used in video action recognition \cite{42,43}, video summarization \cite{44}, and the inference of viscosity from flow simulations \cite{33}. The input to the network consists of processed video frames and the fluid density, while the output is the inferred viscosity (Fig.~\ref{fig:NN}A).

The structure primarily comprises a Convolutional Neural Network (CNN), a Bidirectional Long Short-Term Memory (BLSTM) network, and a temporal self-attention layer, with intermediate pooling, flattening, and normalization layers (Fig.~\ref{fig:NN}A). The CNN is responsible for capturing spatial features within each frame, while the BLSTM captures temporal dependencies across the video, which are crucial for understanding the flow dynamics. Finally, the temporal self-attention layer puts emphasis on the frames based on their relevance. For instance, in lower viscosity cases, most of the fluid movement occurs in the early frames (Fig.~\ref{fig:NN}C(i)), while in higher viscosity cases, flow continues toward the end of the sequence (Fig.~\ref{fig:NN}C(iii)). The self-attention mechanism addresses this by emphasizing frames that contribute most to the viscosity inference. The fluid density input is processed through a dense layer and combined with the video features. The final single-node output provides continuous viscosity inference via regression.

We record 60~s long videos, $t_\text{obs}=60$~s, at 2~fps that we downsampled to 0.5 fps after the first 5~s (Fig.~S1), and we use a $t_\text{flip}=2$~s. We crop the videos within a region of interest and a combination of a Sobel operator with Gaussian Blur filters are used to remove color (Fig.~\ref{fig:NN}C). Videos are augmented by picking 10 different regions of interest (Fig.~\ref{fig:NN}B) through small random translations and rotations of the original region within a few pixels (Fig.~S2). When finding a prediction from a video of a fluid of unknown viscosity, we query the network for all 10 augmentations and report the average and standard deviation from the 10 inferences. 

\subsection*{Training and measurement accuracy}
We collect a dataset of 960 video recordings:\ 10 videos of 96 different viscosity values between 3~mPa$\cdot$s and 1500~Pa$\cdot$s and densities between 0.8~g/mL and 1.4~g/mL (see Materials and Table S1). After augmentation, this yields a total of 9600 videos. The augmentation is always performed after the dataset is split into training and testing sets. We use viscosities obtained from steady shear rotational rheometry as the ground truth (see Methods - Rheometry). By splitting training and test data in two different ways, we evaluate the accuracy of the method in terms of (1) viscosities included in the model training  - aleatoric uncertainty, and (2) viscosities not included in the model training - epistemic uncertainty. 

\begin{figure*}
\centering
\includegraphics[width=16.8cm]{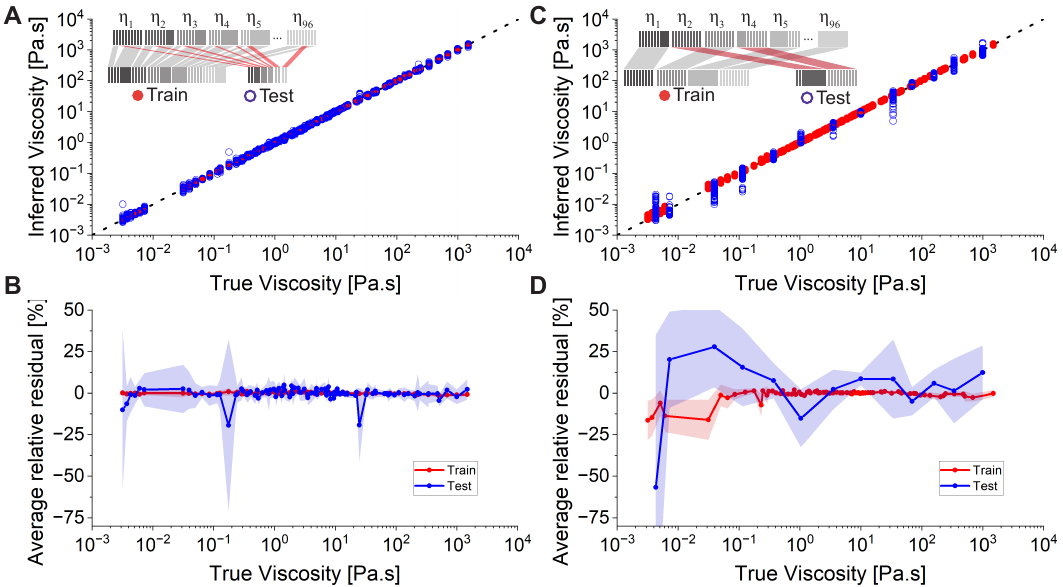}
\caption{Inferred viscosities (A) and relative residuals (B) for fluids with viscosities the model has seen (aleatoric uncertainty). Inferred viscosities (C) and relative residuals (D) for fluids with viscosities the model has not seen (epistemic uncertainty). Shaded regions show the standard deviation of the relative residuals.}
\label{fig:results}
\end{figure*}

To characterize aleatoric uncertainty, we split the data set by randomly picking 8 videos for training and 2 for testing out of the 10 videos of each viscosity (Fig.~\ref{fig:results}A). There is good agreement between true and inferred viscosities. The standard deviations and averages of the residuals generally stay within 25\%, and most often within 5\%, for the entire range tested. Standard deviations of residuals increase close to the upper and lower bounds (Fig.~\ref{fig:results}B). Overall, despite the large viscosity range (almost six orders of magnitude), the inference model exhibits high accuracy, effectively identifying flow features for fluids with viscosities included in its training dataset.

We characterized epistemic uncertainty by splitting the dataset into 13 viscosities for testing and 83 for training (Fig.~\ref{fig:results}C). The test viscosities were purposefully chosen to lie within the overall tested range (interpolation) but in regions with the lowest density of training data points, i.e.\ the most unfavorable conditions for interpolation. In this case, the inferred values exhibit larger average residuals and standard deviations (Fig.~\ref{fig:results}D) compared to the aleatoric accuracy (Fig.~\ref{fig:results}B). This outcome is expected, as the model has not been trained on these specific viscosity values and relies on interpolation for inference. 

Differences in surface tension and contact angles between training and testing fluids can result in epistemic error if these fluid properties significantly affect the flow. Such errors are absent in aleatoric uncertainty evaluations since all test fluids are present in the training set. These effects are most significant at low viscosity ranges where viscous effects may be obfuscated by inertial or capillary effects. This likely explains the larger errors and deviations observed for viscosities below 0.01~mPa$\cdot$s  (see Discussion and Supplementary Information - Effects of surface tension). 

Despite the complex nature of the flow and potential effects from those fluid properties not captured by the model, the epistemic average residuals remain within 25\% for viscosities ranging from 0.01~Pa$\cdot$s to 0.1~Pa$\cdot$s and within 15\% for viscosities above 0.1~Pa$\cdot$s and up to 1000~Pa$\cdot$s. For such a simple method, this represents a high level of accuracy over an extensive dynamic range of viscosity ($\eta$) and flow Reynolds numbers (Re), comparable to that of traditional rheometers and viscometers. This is a remarkable achievement given the complexity of the flow, the simple low-cost equipment, short testing times without sample preparation, and the simple parallelization for high throughput. 

\begin{figure*}[!htbp]
\centering
\includegraphics[width=15.8cm]{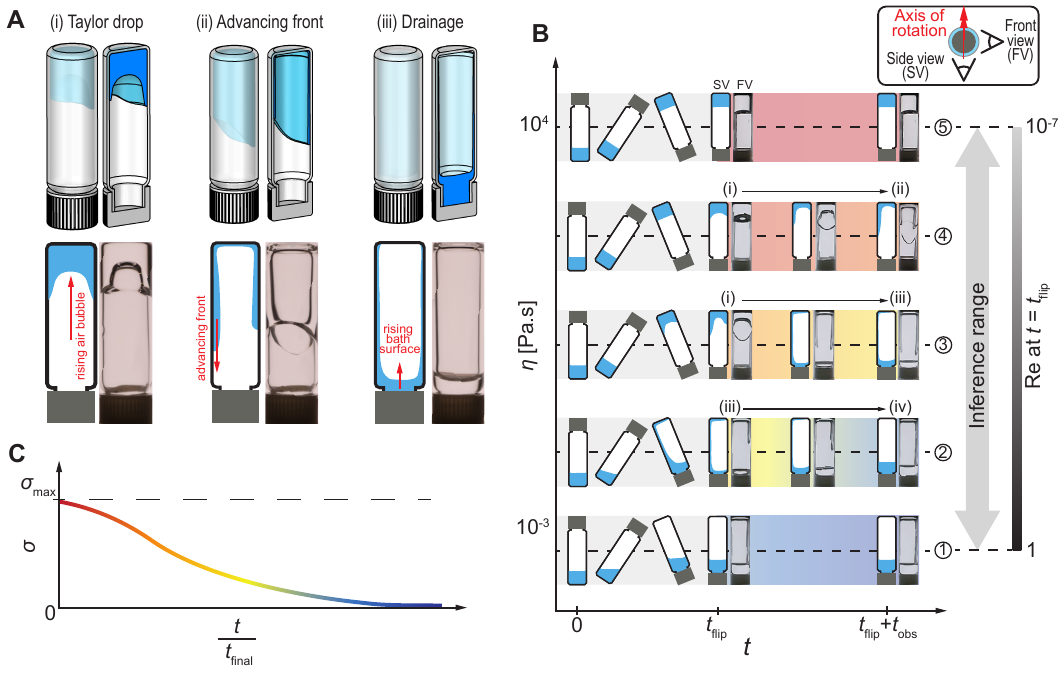}
\caption{
The fluid flow behavior and forcing stress during the inverted vial test.
(A) The flow starts with a Taylor (i) where buoyancy from the trapped air drives the flow. When buoyancy forces become negligible, flow evolves into an (ii) advancing front characterized by a three phase advancing contact line. Once the fluid reaches the bottom, a rising fluid surface defines a (iii) drainage flow. 3D (top) and 2D (bottom-left) simplified schematics and front view pictures (bottom-right).
(B) Flow evolution for finite $t_{\text{flip}}$. Roman numerals indicate flow regimes as shown in (A). The flow transitions into different regimes within $t_{\text{obs}}$, with viscosity strongly influencing the flow regimes and initial conditions.
(C) Evolution of the forcing stress under the assumption of an instantaneous vial flip. The stress starts at the maximum value, $\sigma_\text{max}$, and decreases asymptotically as the flow evolves.
}
\label{fig:realtest}
\end{figure*}

\section*{Discussion}
Our results confirm the feasibility of quantitatively inferring viscosity over five orders of magnitude only from video recordings, without explicit feature tracking or velocity field measurement. The inference is within 25\% error (decreasing to about 15\% above 0.1~Pa$\cdot$s) despite the neural network receiving no velocity‑field data per se, each sample beginning from a distinct flip‑induced initial state, and certain viscosity ranges having finite Re and low Bo flow that may violate classical viscous‑gravity theory. To understand why the network retains accuracy and where it fails, we discuss the complex flow dynamics and how they are affected by process parameters and fluid properties. 

The broad inference range is a direct result of the neural network’s ability to approximate the complex inverse function, accounting for diverse flow scenarios and initial conditions (Fig.~\ref{fig:realtest}A-B). Overall, we observe three distinct flow regimes (Figure \ref{fig:realtest}A) that we refer to as Taylor drop \cite{45,46,47}, advancing front, and drainage \cite{41,48} (Fig.~\ref{fig:realtest}A, and Supplementary Information - Detailed description of the flow and its forcing stresses). The idealized axisymmetric Taylor drop (Fig. \ref{fig:realtest}A(i)) and the final drainage (Fig. \ref{fig:realtest}A(iii)) have been previously analyzed \cite{41,45,46,47,48}. Our tests are not precisely axisymmetric due to the vial rotation, and the transition regime (``advancing front'') comprises a significant amount of the observation and involves the complexity of an advancing contact line, rotational asymmetry, a film that may not be ``thin'', and a complex boundary condition at the top of the vial (Fig. \ref{fig:realtest}A(ii)). Not all regimes are observed for every viscosity. In highly viscous fluids, $\eta \ge 30$~Pa$\cdot$s, we observe that flow remains in the Taylor drop and advancing front regimes within the observation window and may not reach the drainage regime (Fig.~\ref{fig:realtest}B - $\textcircled{4}$). For low to mid-viscosity fluids, significant flow occurs during the flipping process itself, making the initial condition, defined as the moment the vial is fully inverted, inherently viscosity-dependent. For example, low-viscosity fluids, $\eta \le 1$~Pa$\cdot$s, (Fig.~\ref{fig:realtest}B - $\textcircled{2}$) typically enter observation in a drainage-like state, whereas mid-viscosity fluids, $\eta \approx [1-30]$~Pa$\cdot$s, (Fig.~\ref{fig:realtest}B - $\textcircled{3}$) begin from the advancing front regime.

Alternative approaches to using a neural network exist but are fundamentally limited in dynamic range for this flow. For example, one alternative is to restrict inference to flow regimes where approximate solutions are available \cite{46,48,49,50,51}. However, since no single solution captures the full dynamics of the inverted vial test, inference would need to be confined to the extreme cases where the flow approximates either a Taylor drop or a drainage regime. A model based solely on symmetric Taylor drop dynamics might only apply to viscosities above about 30~Pa$\cdot$s, while a symmetric thin-film drainage-based model may require excessive observation times. Another alternative approach is to apply a Buckingham-$\Pi$ analysis. In regimes with negligible inertia (Re$=\rho^2 g R^3/\eta^2\ll$1 based on visco-gravity velocity, $U_{vg}\sim \rho g R^2/\eta$) and surface tension (Bo$=\rho g R^2/\Gamma\gg$1), the characteristic flow velocity scales inversely with viscosity (see Supplementary Information – Calculation of non-dimensional numbers and \cite{12}). However, this approach fails to capture the variable initial condition, which depends on viscosity, and the effects of low Bo and finite Re that pollute the inference. Instead, the neural network covers all regimes, all initial conditions, and can infer for finite Re and low Bo, presumably because some portion of the flow is measurably sensitive to viscosity. The result is an exceptionally broad dynamic range of inference.

The error in inference is minimum (within about 15\%) in the viscosity range spanning from about 0.1 to 1000~Pa$\cdot$s (Fig. \ref{fig:results}D). This is a dynamic range of 10$^4$, two orders of magnitude larger than the temporal and spatial dynamic ranges of the measurement. Within this viscosity window, the low errors suggest that the flow is sensitive to viscosity compared to potential surface tension or internal effects. Also, the characteristic timescales of the flow are on the order of the observation window, resulting in clear observable motion ($>$10 pixel motion within the 60~s observation time). As viscosity shifts outside the 0.1 to 1000~Pa$\cdot$s range, either lower or higher, inference errors increase. 

The limits and errors of the viscosity inference depend critically on $\overline{\text{P}}$ (Eq.~\ref{inverse fucntion}), i.e.\ both operational parameters and other intrinsic fluid properties that we do not consider as input. To investigate how $\overline{\text{P}}$ influences our inference, we conducted a series of additional experiments that include varying the observation window ($t_\text{obs}$) and flip time ($t_\text{flip}$), introducing surfactants, altering optical conditions, and testing non-Newtonian fluids to probe the boundaries of Newtonian assumptions (see Supplementary Information - Additional experiments). Together, these studies define the dynamic range of the method, identify key sources of inference error in the complex flow, outline strategies to mitigate these errors, and reveal opportunities and guidelines for extending its applicability to diverse and complex fluid systems.



To investigate the upper bound of the viscosity range, we test with a different observation window ($t_\text{obs}$). In principle, there is no intrinsic limit on the upper bound of viscosity inference (Fig.~\ref{fig:realtest}B - $\textcircled{5}$), as this limit is determined by the magnitude of observable motion within $t_\text{obs}$. Since all fluids will eventually flow—“panta rhei” (everything flows) \cite{52}—sufficiently long observation should, in theory, produce detectable motion within the spatial resolution of the imaging system, although in practice material mutation~\cite{7,26} may limit the effective observation time. In the high-viscosity regime, where Re is low and Bo is high (Fig.~S3), the total observed displacement over the observation time $t_\text{obs}$ scales as $\Delta x_\text{obs} \propto \rho g R^2 t_\text{obs} / \eta$. Experiments confirm this: inference with a shorter $t_\text{obs}$ down to 10~s, proportionally decreases the observed displacement (about 2 pixels for the most viscous fluids tested), leading to increased inference errors near the upper viscosity limit (Fig.~S4). Increasing vial diameter or spatial resolution may allow for more accurate inference at high viscosities over shorter timescales, which could be particularly valuable for studying highly viscous, reactive systems where test duration must remain below the material mutation timescale~\cite{7,26}.

\begin{figure*}
\centering
\includegraphics[width=16.8cm]{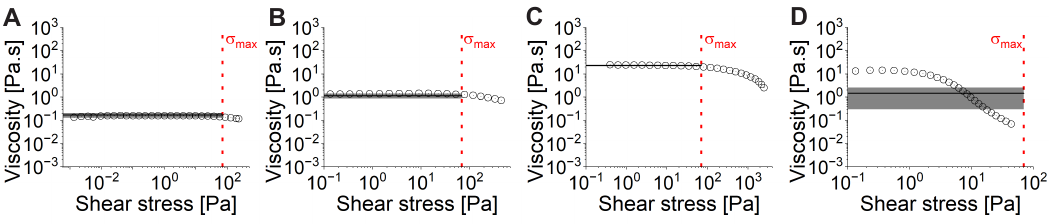}
\caption{Steady shear rheometry (symbols) and inverted vial inference (solid lines) for (A) 2 wt\% 500k polyisobutylene (PIB) in S6, (B) 15 wt\% polyvinylpyrrolidone (PVP) in deionized water, (C) 25 wt\% polyvinylpyrrolidone (PVP) in deionized water, (D) 1 wt\% 8M polyethylene oxide (PEO) in deionized water. Shaded regions show standard deviation of the inverted vial inference. $\sigma_\text{max}$ is the maximum estimated forcing stress during the inverted vial test (see Supplementary Information - Detailed description of the flow and its forcing stresses)}
\label{fig:nonNew}
\end{figure*}

Rationalizing the lower viscosity limit (Fig.~\ref{fig:realtest}B - $\textcircled{1}$) is more complex as it depends on several factors: the visibility of short-time dynamics, fluid inertia obfuscating viscous effects, and surface tension influence on flow timescales. To gain insight into this lower limit, we examined how inference is affected by different flip times ($t_\text{flip}$), which modulate short-time dynamics and inertial contributions, and the addition of surfactants, which alter surface tension.

To study the role of observable early-time flow and inertia, we vary the flip time, $t_\text{flip}$. For long $t_{\text{flip}}$, low-viscosity fluids can equilibrate to follow the flip angle, such that most of the fluid settles at the bottom by the time the flip is complete, leaving only a thin film behind. As a result, limited motion is observed in the first few frames (Fig.~S5A-C), which limits the amount of flow information available for inference. Shorter $t_{\text{flip}}$ may result in less flow during the flip but risk chaotic sloshing motion for low viscosities. In the absence of sloshing, shorter $t_{\text{flip}}$ creates an initial condition with more fluid remaining at the top, allowing for more observable flow during $t_\text{obs}$ (Fig.~S5E,F). This leads to lower absolute errors and reduced standard deviation in viscosity inference at low viscosities, up to a threshold beyond which error suddenly increases (Fig.~S5G). Below this threshold, inertial effects associated with sudden acceleration-deceleration during the flip (characterized by Re$_{\text{flip}}$, see Supplementary Information - Calculation of non-dimensional numbers) introduce wave-like flow dynamics (Fig.~S5D) that increase the initial Re by an order of magnitude in the lower viscosity limit (Fig.~S3) and seem to obfuscate viscosity-dependent behavior. These findings suggest that optimizing $t_{\text{flip}}$, to maximize observable flow without introducing strong inertial effects from the flip acceleration, may improve inference in the low-viscosity regime. 

To study the effects of surface tension, we test three low viscosity fluids ($\eta\approx 0.01, 0.1, 1$~Pa$\cdot$s) with added surfactant to reduce their surface tension (see Supplementary Information - Effects of surface tension). Our results suggest that reducing surface tension leads to viscosity overestimation of up to 40\% for fluids with $\eta\approx10$~mPa$\cdot$s, decreasing to 20\% error for $\eta\approx1$~Pa$\cdot$s (Fig.~S6). In the low-viscosity range, potential low Bond numbers during drainage result in surface tension effects that are non-negligible, reducing the characteristic timescale compared to purely gravity-driven flow and thereby leading to viscosity overestimation (see Supplementary Information – Effects of Surface Tension). Since the current inverse function (Eq.\ref{inverse fucntion}) accounts only for fluid density and does not explicitly incorporate other fluid properties such as surface tension or contact angle, systematic errors can arise when fluids differ in these later properties. Including surface tension and contact angle as additional input features may help mitigate these errors and improve robustness across a broader range of fluids.

Because inference relies on the raw videos, optical parameters such as frame rate and fluid color can also influence accuracy. To understand the impact of frame rate, we retrained and tested the model using only the first and last frames. Results showed minimal effects on mean accuracy but increased deviation (Fig.~S7). This suggests that initial and final conditions are sensitive to viscosity, and additional frames refine feature extraction. Fluid opacity and color can also pose challenges. To investigate this, we tested the current model, trained mostly on clear fluids, with opaque or colored fluids. Results showed significantly higher error, particularly at viscosity extremes where observable flow is limited (Fig.~S8 - red). Retraining the model with colored samples experimentally demonstrated that expanding the training set can mitigate this effect (Fig.~S8, blue). Further exploration of image processing techniques may also help address this challenge. However, completely opaque fluids that coat the inner surface of the vial, e.g.\ paints, may not be suitable for this method, as they obscure optical features of the flow. In such cases, the advancing front may still be visible, but not the drainage regime, thus limiting the inference.

While the model is based on Newtonian assumptions (Eq.~\ref{inverse fucntion}), we intentionally tested it using non-Newtonian fluids to evaluate how deviations from these assumptions affect inference. We hypothesize that reliable inference remains possible when two conditions are met: (1) viscoelasticity is negligible, i.e.\ the Deborah number $\mathrm{De} = \tau / t_\text{flow}$ is much less than 1, and (2) nonlinearities are negligible, i.e.\ the dimensionless stress amplitude $\mathcal{A} = \sigma_\text{flow} / \sigma_{\text{crit}}$ is also much less than 1. Here $\tau$ is the characteristic viscoelastic relaxation timescale of the material and $\sigma_\text{crit}$ is the critical stress for the material to deviate from linear viscoelasticity~\cite{53}; $t_\text{flow}$ and $\sigma_\text{flow}$ are the timescale and stress associated with the observed flow condition~\cite{12,17,54}.

To evaluate this hypothesis, we first approximate the characteristic stress and timescale associated with the test (this is also good practice to avoid erroneous qualitative inference \cite{12,17}). The stress evolves during the test, but it is upper bounded and thus, $\sigma_{\text{flow}}\le ~ \sigma_\text{max}= \Delta \rho g R\approx70$~Pa (Fig.~\ref{fig:realtest}C and Eq.~S1). The relevant timescale for inference is constrained between two bounds. The lower bound is set by the transient flip time, $t_{\text{flip}}$, since flow dynamics during the flip are not captured. The upper bound is the observation cutoff time, $t_{\text{obs}}$, beyond which no additional data is collected. Thus, the characteristic flow time lies within the range $t_{\text{flip}}<t_\text{flow}<t_{\text{obs}}$.

We tested four shear-thinning viscoelastic polymeric solutions (see Materials) with different zero-shear viscosities and shear-thinning behaviors (Fig.~\ref{fig:nonNew}) to assess the reliability of our inference in the presence of non-Newtonian behavior. To estimate De and $\mathcal{A}$, we characterize the viscoelastic and shear-thinning behavior of the fluids. 

Linear viscoelastic analysis (Fig.~S9) reveals relaxation times of $\tau=$ 0.005, 0.007, 0.2, and 4~s. While the tilted vial test is a stress-driven flow and more accurately characterized by a retardation timescale ($\tau_r$), the relaxation timescale $\tau$ provides a conservative upper bound, as $\tau_r<\tau$~\cite{55}. This estimate is also consistent with the upper bounds obtained from creep experiments (Fig.~S10). The De numbers for these flows are therefore upper bounded by De~=~0.003, 0.004, 0.1, and 2 (from $t_\text{flow}=t_\text{flip}$). 

Critical stresses were obtained from fitting an Ellis model \cite{56} (see Fig.~S11 and Table~S2) of the form $\eta(\sigma)=\eta_0\left(1+(k\sigma)^{a-1}\right)^{-1}$ where $\eta_0$, $k$ and $a$ are fitting parameters and $\sigma_{crit}=1/k=5.8\cdot 10^8$, 526, 476, and 3.7~Pa is related to a critical stress for nonlinearity to steady-shear viscosity. Using $\sigma_{\text{flow}}=70$~Pa, gives $\mathcal{A}=~\sigma_\text{max}/\sigma_{\text{crit}}=$ 10$^{-7}$, 0.13, 0.15, and 19.  These are maximum values of $\mathcal{A}$ that decrease as the characteristic stress $\sigma_{\text{flow}}$ decreases from $\sigma_\text{max}$ during the test (Fig. \ref{fig:realtest}C and Supplemental Information - Detailed description of the flow and its forcing stresses).

The CV viscometer infers viscosity with high accuracy and small standard deviation for flows with De$<1$ and $\mathcal{A}<1$ (Fig.~\ref{fig:nonNew}A-C). Essentially, the inference is accurate if the non-Newtonian fluid behaves as Newtonian within the estimated stress range and timescales of the inverted vial test. This observation confirms that our $\sigma_{\text{max}}$ estimate is within acceptable bounds.

For the fluid with $\text{De}=2$ and $\mathcal{A}=19$ (Fig.~\ref{fig:nonNew}D), the CV viscometer infers a viscosity value that approximates an average over the stress range up to $\sigma_{\text{max}}$ with significantly larger standard deviation compared to the other fluids. To gain further insight, we compare the concatenated frames of this non-Newtonian fluid with those of Newtonian fluids (Fig.~S12). The initial non-Newtonian fluid frames resemble those of a 1~Pa$\cdot$s Newtonian fluid (regime of max $\sigma_\text{flow}$ and $\mathcal{A} > 1$), while later frames appear similar to those of a 5~Pa$\cdot$s Newtonian fluid (as $\sigma_\text{flow}$ decreases over time and $\mathcal{A} < 1$). The different regions of interest used for inference (see Methods - Querying the viscosity of non-Newtonian fluids) likely emphasize some time instants more than others, resulting in a high standard deviation. While the model does not yet characterize full non-Newtonian behavior, the presence of large standard deviations in the inferred viscosity can be used as a potential proxy of shear-thinning behavior.

\section*{Conclusion}
We demonstrate that fluid viscosity can be inferred from visual features of complex flows using a neural network that approximates the inverse problem directly from video input, without requiring explicit velocity field measurements. Based on this principle, we developed the CV viscometer, a high-throughput and accessible system for viscosity inference. Our results show that neural networks can solve inverse problems of complex flows from video recordings - a remarkable finding given that video recordings capture optical features rather than precise velocity field measurements, yet the model successfully extracts viscosity-relevant information.

The CV viscometer and the neural network infer viscosity ranges comparable to those of typical viscometers despite the simplicity of the method. Unlike extrusion-based~\cite{38, 57} or microfluidic-based~\cite{10} viscometers, the CV viscometer performs parallel measurements using a single, simple sensor: a camera. It eliminates the need for sample transfer or cleaning and can integrate into automated or manual laboratories that use vials, offering a highly accessible high throughput testing method. 

Of course, its simplicity, accessibility, and high throughput come with trade-offs. Traditional methods typically achieve higher accuracies (1–3\%) and can characterize non-Newtonian behavior. The CV viscometer primarily provides an estimate of constant $\eta$ (with 15-25\% accuracy) while detecting non-Newtonian behavior from inference uncertainty. While this may seem modest, it is remarkably good when applied across orders of magnitude, as these errors translate to minor shifts on a log scale, making the method effective for its five orders of magnitude dynamic range. In addition, despite its higher error compared to traditional methods, the CV viscometer’s accuracy satisfies requirements for high-throughput material screening and discovery~\cite{2,3,4,26}, where rapid, approximate measurements are often more valuable than time-intensive, precise ones for selecting promising candidates. Further model development and additional training may improve non-Newtonian characterization, and the gravity-driven nature of the test offers opportunities to adjust forcing stresses through geometric design.

Beyond enabling viscosity inference, the CV viscometer highlights the potential of rethinking conventional materials testing to develop faster and more accessible characterization techniques. It exemplifies how machine learning can harness the growing accessibility of imaging and computing power, even in our pockets, to advance scientific methodologies. Similar automation approaches applied to other protorheology tests~\cite{12} could extend this concept to infer additional rheological quantities, such as viscoelastic properties, extensional viscosities, and normal stress differences, among others. Although our study focuses on a systematic setup for vial inversion, we envision future developments where neural networks infer rheological properties from manually actuated protorheology tests captured by handheld devices. The accessibility of this method enables high throughput screening for material discovery in automated laboratories and makes rheological characterization more broadly available through the use of simpler and more affordable tools.

\subsection*{Materials}

\subsection*{Training/ testing (Newtonian) fluids}
For training and testing of the neural network, we use 16 fluids: two S (Cannon, USA), five N (Cannon, USA), three RMT (Sigma Aldrich) and one RT (Sigma Aldrich) viscosity standards made of highly-refined oil (Table~S1), glycerol (Sigma Aldrich), raw unfiltered honey (Nate's Honey Co.), corn syrup 1 (Fischer Science), and  corn syrup 2 (Lab Alley Essential Chemicals). We test fluids at 6 different temperatures resulting in 96 viscosities. We report densities and viscosities of all fluids in Table S1.

\subsection*{Non-Newtonian fluids}
15wt\% and 25wt\% polyvinylpyrrolidone (Sigma Aldrich, $M_n\approx 360$ kDa) solutions were prepared by dissolving the polymer in 80°C deionized water and then cooling to room temperature. 1wt\% polyethylene oxide solution (Sigma Aldrich, $M_w \approx 8,000,000$) were made by dissolving PEO powder in a 19:1 de-ionized water–ethanol co-solvent  by vigorously mixing in a beaker with a magnetic stir rod and stir plate at 200 rpm for 12 h until a homogeneous viscoelastic solution was obtained. 2wt\% polyisobutylene (Sigma Aldrich, $M_w \approx 500,000$) were prepared by dissolving PIB powder in S6 (Canon, USA) at 70°C at 300 rpm for 6 h. 

\subsection*{Methods}

\subsection*{CV viscometer}
The CV viscometer consists of a stepper motor that connects to a custom-made vial holder through a flexible coupling. It holds six vials at a time and has a rubber gasket at the bottom that prevents the vials from rotating around their axes as they are inverted. A web camera is placed about 200 mm away from the center of rotation of the vial holder. A software written in Python language actuates the stepper motor through an Arduino microcontroller and a TXB6600 stepper driver set to 1600 steps per revolution. The Arduino is actuated from the Python code using the serial port and the AccelStepper library to run the stepper motor and control maximum accelerations and velocities. The AccelStepper library parameters are chosen such that a 180$^\text{o}$ inversion is obtained in 2 seconds unless otherwise noted. The same Python software starts the video recording before the inversion of the vial and records for a total of 60 s. The camera is set to record at 2 frames per second with a total resolution of 1280 x 720 pixels. The video is then separated into the 60 different regions of interest, 10 for each vial, of 104 by 10 pixels and sent as input to the neural network for training and testing. 

\subsection*{Model training}
For training, we place 2 mL of each training fluid in an 8 mL vial (see Materials) using a syringe. We place the CV viscometer inside an environmental chamber and flip the vials at environmental temperatures equal to 25, 30, 35, 40, 45, and 50°C. The fluids are first randomly placed in the six CV viscometer positions. Then, they are flipped 10 times with 10 minute intervals between flips. We fill the six positions of the CV viscometer as much as possible without repeating fluids to reduce the number of tests. 

Videos are separated into training and test data sets depending on the accuracy to be queried. The videos are processed and augmented 10 times according to the main text. For training, we use an Adam optimizer with an initial learning rate equal to 1e-5 and the mean squared error as the loss function. We run the training for 2000 epochs or until the loss does not improve after 1000 epochs. We use measurement from parallel plate rheometry (see Methods-Rheology) and measured density (see Methods-Density measurements) as the ground truth.

\subsection*{Querying the viscosity of non-Newtonian fluids}
For the non-Newtonian fluids, 2 mL of each of the fluids is transferred to a 8 mL vial using a syringe. The vials are then randomly placed in the different positions of the CV viscometer holder. The viscometer software was run to flip the vial and record a single video at 23°C. The video was then run through optical filters and augmented 10 times. The 10 augmentations, together with the fluid density, are then input into the machine learning model, resulting in 10 viscosity outputs. The inference is the average of the 10 viscosities and the standard deviation.

\subsection*{Density measurement}
The densities of the training fluids are taken from the certificate analysis provided by the manufacturer if available and linearly interpolated to temperatures not provided. For fluids where densities are not provided by the manufacturer, an Anton Parr DMA-501 density meter is used for density measurement. Density values for temperatures not accessible by the equipment are calculated by linear interpolation.  Fluids N1900000, N4500000, and CornSyrup2 are too viscous to be measured using the density meter, thus, we approximate the densities of these two N standard fluids with those of the N standard with the closest viscosity, i.e.\ N62000, and those of CornSyrup2 with the ones from CornSyrup1. Table S1 reports the densities of all training fluids.

\subsection*{Rheometry}
The viscosity of the training fluids are measured using an ARES-G2 rotational rheometer (TA Instruments) with cone-and-plate geometries of 25 mm  ($\beta= 5.73^{\text{o}}$) and 50 mm ($\beta = 1.994^{\text{o}}$). Viscosity was determined over a temperature range of 20°C to 60°C using a flow temperature sweep protocol with a ramp rate of 1°C/min at a shear rate of $\dot\gamma=10$~s$^{-1}$. Temperature control was achieved using the Advanced Peltier System (APS) and data were collected with a sampling interval of 10 s/point. Viscosity values are provided in Table S1. Steady shear data for different polymer solutions were obtained using a 40 mm parallel plate geometry (DHR-3, TA Instruments, including corrections for parallel plate ) with a 1 mm gap at 25°C (Fig.~\ref{fig:nonNew}).


\section*{Acknowledgment}
This research was conducted in the Autonomous Materials Systems group at the Beckman Institute for Advanced Science and Technology at the University of Illinois Urbana-Champaign. This work was supported as part of the Regenerative EnergyEfficient Manufacturing of Thermoset Polymeric Materials (REMAT), an Energy Frontier Research Center funded by the U.S. Department of Energy, Office of Science, Basic Energy Sciences under award DE-SC0023457.




\end{document}


\title{Supplementary information for high-throughput~viscometry~via machine-learning from~videos~of inverted vials}

\author[a,1,2]{Ignacio Arretche$^{\text{a,1,2}}$, Mohammad Tanver Hossain$^{\text{a,b,1}}$, Ramdas Tiwari$^{\text{a,b}}$, Abbie Kim$^{\text{a,b}}$, Mya G. Mills$^{\text{c}}$, Connor D. Armstrong$^{\text{a,d}}$, Jacob J. Lessard$^{\text{e}}$, Sameh H. Tawfick$^{\text{a,b,2}}$, and Randy H. Ewoldt$^{\text{a,b,2}}$ }
\date{}  
\maketitle

\noindent$^{\text{a}}$Beckman Institute for Advanced Science and Technology, UIUC, Illinois 61801\\
$^{\text{b}}$Department of Mechanical Science and Engineering, UIUC, Illinois 61801\\
$^{\text{c}}$Department of Chemistry, UIUC, Urbana, Illinois 61801\\
$^{\text{d}}$Department of Material Science and Engineering, UIUC, Urbana, Illinois 61801\\
$^{\text{e}}$Department of Chemistry, University of Utah, Salt Lake City, Utah 84112\\
$^{\text{1}}$ IA and MTH contributed equally\\
$^{\text{2}}$To whom correspondence should be addressed.\\ E-mail: ia6@illinois.edu, tawfick@illinois.edu and ewoldt@illinois.edu


\maketitle

\subsection*{Detailed description of the flow and its forcing stresses}
We describe each of the three regimes (Fig.~5A) of the complex flow of the inverted vial test and estimate the associated forcing stresses, both of which are critical for understanding the accuracy of the method, why it fails, and when it is affected by non-Newtonian behavior (see Discussion). Assuming the flip time is ``instantaneous'' (i.e., initial viscosity independent conditions), the flow starts with a Taylor drop, characterized by an air bubble rising into the viscous fluid. Although we observe a slight asymmetry from flipping, the flow here is approximately asymmetric (Fig. 5A(i)). The forcing stress during the Taylor drop regime is driven by buoyancy, and it can be estimated as~\cite{1}
\begin{equation}
    \sigma_{\text{Taylor}}\approx\Delta \rho g R
    \label{eq:charact_1}
\end{equation}
where $\Delta \rho$ is the difference in density between the fluid in the vial and air, and $R$ is the interior radius of the vial. Assuming that the density of air is negligible compared to that of the test fluids, $\Delta \rho = \rho$ , where $\rho$ is the density of the test fluids. Taking an estimated nominal density of the fluids to be 900 kg/m$^{\text{3}}$~(Table \ref{Table:Table}) and $R=$7.5 mm, the resulting stress is $\sigma_\text{Taylor} \approx 70$ Pa. 

As the test progresses, the air bubble approximates the top of the vial, and stresses from buoyancy forces become negligible. An advancing front (Fig. 5A(ii)) characterized by a three-phase advancing contact line forms down the sides of the vial. The flow here is strongly asymmetric; the film flow may not always be ``thin'', and the boundary condition at the top is complex and hard to determine. Due to the complexity of the flow and the lack of approximate solutions, determining the forcing stresses of this flow requires further numerical studies, which we leave for future work. For now, we hypothesize that the forcing stress of the advancing front monotonically transitions to the next flow regime, which we describe in more detail in the next paragraph. Finally, we note that although fingering instabilities may be present for similar flows~\cite{2,3}, we do not observe any here. 

As the advancing front evolves, it reaches the bottom, and the vial starts to fill up. Here, the flow transitions into drainage (Fig. 5A(iii))~\cite{4,5,6,7}, i.e.\ a film flow with an upward-moving bath surface. Drainage-like flows \cite{4,5,7,9} have been studied in the literature, and approximate solutions to these are available far from the boundaries where (1) the effect of the hydrostatic pressure gradient is much smaller than the direct action of gravity (this may be violated at the top of the vial), and (2) effects of surface tension are negligible (this may be violated near the bath surface, where film thickness suddenly increases to $R$, resulting in a significant free surface curvature). 

Assuming no-slip boundary conditions and $R>>h$, where $h$ is the film thickness, the velocity profile for the drainage of a Newtonian fluid takes a parabolic form \cite{8,9,2,7,4}
\begin{equation}
    u(x,z,t) = \frac{\rho g}{\eta} \left(h(x,t) z-\frac{1}{2}z^2\right)
    \label{eq:charact_2}
\end{equation}
where $z$ is the distance normal to the vial surface, $x$ is the vertical distance from the top of the vial, and $t$ is the time. In this case, the shear stress is given by
\begin{equation}
    \sigma(x,z,t) = \eta \frac{\partial u}{\partial z} = \rho g (h(x,t)-z),
    \label{eq:charact_3}
\end{equation}
and the film thickness evolves as \cite{9,7,2}
\begin{equation}
    h(x,t) = \sqrt{\frac{\eta x}{gt}}
    \label{eq:charact_4}
\end{equation}
Combining Eqs. \ref{eq:charact_3} and \ref{eq:charact_4}, the characteristic stress as a function of time during drainage flow is 
\begin{equation}
    \sigma(x,z,t) = \rho g \left( \sqrt{\frac{\eta x}{gt}} -z \right).
    \label{eq:charact_5}
\end{equation}

In drainage, the forcing stress scales with $\sigma\propto t^{-1/2}$. We hypothesize that the stress during the inverted vial test starts at maximum value $\sigma_\text{max}=\sigma_\text{Taylor}\approx70$~Pa (Eq.~S\ref{eq:charact_1}), then monotonically decreases through the advancing front regime, and finally asymptomatically approaches zero during drainage (Fig.~5A and \ref{fig:flowevol}A). Because the stress decreases from $\sigma_\text{max}$ throughout the test, the fluid is not subjected to a single stress value but rather to a range of stresses. Although further numerical analysis is needed to confirm the details of this stress evolution, our inference results for non-Newtonian fluids provide initial evidence that the estimated stresses are reasonably accurate (Fig.~6).


\subsection*{Estimation of non-dimensional numbers}
Since no working equations are previously available for the inverted vial test, the determination of the characteristic and non-dimensional quantities for the test is not readily available. A Buckingham~$\Pi$ analysis to obtain a characteristic velocity ($U$) associated with the test \cite{8}, can be initiated by postulating the functional dependence as
\begin{equation}
    U=f(\eta, \rho, \rho g, H, R, \theta, \Gamma, \beta),
    \label{eq:1_SI}
\end{equation}
where $H$ is the liquid height, and $\beta$ is the contact angle. Buckingham-$\Pi$ analysis results in six $\Pi$ terms, hence, the characteristic velocity associated with the viscous-gravity flow can be written as
\begin{equation}
    U=\frac{\rho gR^2}{\eta} \phi \left(\frac{H}{R}, \theta, \frac{\rho g R^2}{\Gamma}, \frac{\rho (\rho g) R^3}{\eta^2}, \beta \right)
    \label{eq:2_SI}
\end{equation}
where $\rho g R^2/ \Gamma$ is the Bond number and $\rho (\rho g) R^3/ \eta^2$ can be written as a visco-gravity Reynolds number,
\begin{equation}
    \text{Re}_{\text{VG}}= \frac{\rho (\rho g) R^3}{\eta^2} = \frac{\rho R }{\eta}\cdot \frac{\rho g R^2}{\eta}=\frac{\rho R }{\eta}\cdot U_{VG}.
    \label{eq:ReVG}
\end{equation}
The term $U_{VG}=\rho g R^2/ \eta$ is a visco-gravity velocity ($U_{VG}$) arising from balancing the viscous stresses $\eta U_{VG} / R$ and gravitational stress $\rho g R$.
Neglecting inertia (Re~$\ll1$), surface tension (Bo~$\gg1$), and the effects from contact angles results in a $U$ governed by the fluid viscosity as
\begin{equation}
    U=\frac{\rho gR^2}{\eta} \phi \left(\frac{H}{R}, \theta \right).
    \label{eq:4_SI}
\end{equation}




The Buckingham $\Pi$ analysis is adequate as long as the initial conditions of the test do not change, i.e., they are independent of fluid viscosity. However, for fluids with viscosities $\eta\lesssim 30$~Pa$\cdot$s, we observe that the initial condition (defined as the instant when the vial is fully inverted) changes from a Taylor drop to either an advancing front (1~Pa$\cdot$s~$\lesssim\eta\lesssim 30$~Pa$\cdot$s) or drainage ($\eta\lesssim 1$ Pa.s), making this analysis invalid. 

To experimentally validate the non-dimensional Re$_\text{VG}$ expression from Buckingham-$\Pi$ (Eq.~S\ref{eq:ReVG}), we optically measure the velocity of the fluid free surface to estimate an experimental Re$_\text{exp}$. This Re$_\text{exp}$ is only an approximation, as the fluid free surface may be moving at different speeds than the bulk of the fluid. We calculate the Re$_\text{exp}$ at the initial condition.

Since the flow shows three different fluid regimes, we must define the length scale and velocity for each regime. In the case of an initial Taylor drop, we optically measure the initial velocity of the meniscus $U_\text{men}$ and calculate Re$_\text{exp}$ as
\begin{equation}
    \text{Re}_\text{exp}= \frac{\rho U_\text{men} R}{\eta}
\end{equation}
In the case of drainage, we define Re$_\text{exp}$ as
\begin{equation}
    \text{Re}_\text{exp}=\frac{\rho U_\text{rise}R}{\eta}
    \label{eq:Re_drainage}
\end{equation}
where $U_\text{rise}$ is the velocity of the rising bath surface. Using conservation of mass, $U_\text{rise}\pi R^2 \approx U_\text{film} 2\pi R h$, which leads to $\rho U_\text{rise} R / \eta = 2\pi U_\text{film} h / \eta$, where $U_\text{film}$ and $h$ are the velocity and thickness of the thin downward film, respectively. Essentially, Reynolds numbers measured from film velocities and length scales are of the same order as those measured from the rising bath velocities and length scales. Since we cannot directly measure $U_\text{film}$ and $h$, but we can measure $U_\text{rise}$ and $R$, we adopt Eq.~S\ref{eq:Re_drainage} as the definition of Re$_\text{exp}$ for drainage.

In the case of the advancing front, the Re$_\text{exp}$ is more challenging to define given the asymmetry complexity of the flow, and our inability to measure the front thickness, i.e. the characteristic length scale associated with this flow. We choose to estimate it as
\begin{equation}
    \text{Re}_\text{exp}=\frac{\rho U_\text{front}R}{\eta}
\end{equation}
where $U_\text{front}$ is the velocity of the lowest point of the advancing front. This is a rough approximation, and future numerical solutions for this flow may allow for better estimates. 

Results for Re$_\text{exp}$ as a function of viscosity are shown in Fig. \ref{fig:flowevol}B. The scaling law predicted by Eq.~S\ref{eq:ReVG} shows good agreement with Re$_\text{exp}$ in the Taylor drop regime. Here, initial conditions are independent of viscosity because the timescales of flow are large compared to the flip time and Re$_\text{VG}$ from Buckingham $\Pi$ analysis provides a good estimation of the initial flow conditions. 

As we move toward lower viscosities, Re$_\text{exp}$ deviates from the prediction of Eq.~S\ref{eq:ReVG}. At these lower viscosities, we observe flow during the flip, and by the time the vial is fully inverted, the flow is either in the advancing front or drainage regimes. The values of Re$_\text{exp}$ are lower than those estimated by Eq.~S\ref{eq:ReVG} because the measured velocities are lower than the $U_{VG}$ predicted from Buckingham-$\Pi$ analysis. Essentially, the flow velocities, and thus Re, are ``self-regulated'' by the flip times. While the inherent viscosity-dependence of initial conditions complicates the analysis and invalidates simple scaling solutions such as those from Buckingham-$\Pi$ analysis, we hypothesize the ``self-regulation'' enables inference at low viscosities without significant obfuscation from inertial effects. The ability of the neural network to approximate the inverse function under these different initial conditions, combined with the ``self-regulation'' of the inertial effects, allows for the extremely extensive dynamic range of inference. 

To gain further insights into how flip time $t_\text{flip}$ affects the inference, we also measure Re$_\text{exp}$ for the shorter $t_\text{flip}$, i.e. $t_\text{flip}=0.5$~s. As expected, Re$_\text{exp}$ are unaffected by $t_\text{flip}$ when the flow shows an initial Taylor drop regime. On the contrary, in the low viscosity regime, the shorter $t_\text{flip}$ significantly increases Re$_\text{exp}$. At 4~mPa$\cdot$s, Re$\approx$10 - an order of magnitude higher compared to $t_\text{flip}=2$~s. This results in the higher errors of inference at these low viscosities and short flip times (Fig.~\ref{fig:flipspeed}). 

We can also define a Reynolds number associated with the flip motion itself, considering fluid motion in the tangential direction. While the Re discussed thus far describes axial fluid motion inside the vial, tangential motion during the flip can also induce inertial effects. Faster flips result in higher tangential velocities and accelerations, which can impact viscosity inference, particularly in the lower viscosity regime. We characterize this by

\begin{equation}
    \text{Re}_{\text{flip}} = \frac{\rho R U_{\text{flip}}}{\eta} \approx\frac{\rho R}{\eta} \cdot \frac{R} {t_{\text{flip}}}
       \label{eq:Reflip}
\end{equation}
with a characteristic velocity during the flip given by $U_{\text{flip}} \sim R /t_{\text{flip}}$.
For a flip time of $t_{\text{flip}} = 2$ sec, Re$_{\text{flip}}$ takes values between $\sim 25$ and $\sim 0.025$ for $\eta = 1$ mPa.s and $\eta = 1$ Pa.s, respectively. When $t_{\text{flip}}$ is decreased to 0.5 s, Re$_{\text{flip}}$ increases to $\sim 100$ for $\eta = 1$ mPa.s. In this regime, inertial effects appear as wave-like features in the first few video frames (Fig. \ref{fig:flipspeed}C), which in turn increase the Re$_\text{exp}$ (Fig.~\ref{fig:flowevol}B).

Surface tension may also affect the inference. Since our inverse function (Eq. 4) does not have surface tension as an input, if surface tension significantly affects the flow, differences in surface tension among fluids can induce error. We can characterize the significance of surface tension through the Bond number (Bo$=\rho g R^2/\Gamma$; Eq.S~\ref{eq:2_SI}), which relates the gravitational stress to the capillary stress.  The surface tension for the fluids tested is in the range of [30-50] N/m, which estimates the initial Bo (at $t=t_{\text{flip}}$) to be $\sim [10-17]$ under ``instantaneous'' flip assumptions. However, we estimate gravitational stresses in drainage to be smaller than those of the Taylor drop (Fig.~\ref{fig:flowevol}A), thus we expect smaller initial Bo when initial conditions are in drainage, i.e. at the low viscosities. As a result, surface tension may become non-negligible, especially at low viscosities, as capillary stresses may be significant compared to gravitational stresses. We empirically test this in the Supplementary Information - Effects of surface tension. 


\subsection*{Additional experiments}
To understand the effects of $\overline{\text{P}}$ on the inference, we perform the following experiments:
\begin{enumerate}
    \item Effects of $t_\text{obs}$: reduced $t_\text{obs}$ from 60 s to 10 s (see Fig.~\ref{fig:videolength} and Effects of frame rate and video length). 
    \item Effects of $t_\text{flip}$: reduced $t_\text{flip}$ from 2 s to 0.5 s (see Fig.~\ref{fig:flipspeed} and Effects of flip speed).
    \item Effects of surface tension: inference on fluids S6 ($\eta=0.01$), S60 and N600 with added <1wt\% Span 80 surfactant (see Fig.~\ref{fig:surf} and Effects of surface tension).
    \item Effects of frame rate: reduced frame rate from 2 fps (first 5 seconds) and 1 fps (5 - 60 s) to 0.02 fps (keep only first and last frames)  (see Fig.~\ref{fig:framerate} and Effects of frame rate and video length). 
    \item Effects of fluid opacity: Infer viscosity of fluids with $2.5$ mg/mL of blue dye (see Fig.~\ref{fig:color} and Effects of fluid opacity). 
    \item Effects of non-Newtonian behavior: Infer viscosity of four polymer solutions: 2\%wt 500k polyisobutylene (PIB) in S6, 15\%wt polyvinylpyrrolidone (PVP) in deionized water, 25\%wt polyvinylpyrrolidone (PVP) in deionized water, and 1\%wt 8M polyethylene oxide (PEO) in deionized water (see Figs.~6 and \ref{fig:LVE}-\ref{fig:nonNew} and Discussion).
\end{enumerate}

\subsection*{Effects of frame rate and video length}
To query the effects of frame rate and video length on epistemic accuracy, we use the same videos as those on the main text and post-process them to either only keep first and last frames (lower frame rate - Fig. \ref{fig:framerate}B) or the first 10 seconds (lower observation time - Fig. \ref{fig:videolength}B). The post-processed videos are then divided into the same training and testing sets as those of the main text and the neural network is trained and tested on these videos. Results are shown in Figs. \ref{fig:framerate}C-D and \ref{fig:videolength}C-D, receptively. Lower frame rates result in similar average residuals but larger standard deviations (Fig. \ref{fig:framerate}D). Shorter video lengths increase the error at high viscosities (Fig. \ref{fig:videolength}D). For these shorter videos, high viscosities show small observable flow during the time of inference (Fig. \ref{fig:videolength}B(iii)).

\subsection*{Effects of flip speed}
To query the effects of $t_{flip}$, we run flip tests with a $t_{flip}=0.5$ s, which is the lowest $t_{flip}$ possible with our setup. We train and test the model following the same procedure as for the main text model (see Methods - Model training) and dividing into the same training and testing sets as those of the main text. The $t_{flip}$ has a significant effect at the lower viscosity range (Fig. \ref{fig:flipspeed}G). As $t_{flip}$ is decreased, we observe a decrease in epistemic error for viscosities between $\sim 10$ mPa.s and $ 100$~mPa.s. An increase in flip speed results in more observable flow within the time of inference for these low viscosities (Figs. \ref{fig:flipspeed}B-C and Figs.~\ref{fig:flipspeed}E-F). However, at even lower viscosities (lower than $\sim 10$ mPa.s), the error suddenly increases (Fig. \ref{fig:flipspeed}G). Fig.~\ref{fig:flipspeed}D shows two different videos of the fluid with a viscosity of $4$ mPa.s. The flow within the first few frames shows wave-like features, with appreciable differences between the two videos. The high $\text{Re}_\text{flip}$ (Eq.\ref{eq:Reflip}) induces wave-like flow that considerably increases $\text{Re}_\text{exp}$ (Fig.~\ref{fig:flowevol}B), suggesting that the high error arises from obfuscation by inertial effects.

\subsection*{Effects of surface tension}
 To query the effects of surface tension on the inference, we add a surfactant (Span 80, <1wt\%) to the 3 standard fluids of lowest viscosity (S6, S60 and N600 fluids) to lower their surface tension (Fig.~\ref{fig:surf}). We then infer the viscosity of the modified fluids at room temperature using the CV viscometer (Fig.~\ref{fig:surf}). The apparent inferred viscosity for the fluids with surfactant is higher compared to those without surfactant (Fig.~\ref{fig:surf}B), implying surface tension aids the flow. Lower surface tension results in longer characteristic times and thus higher apparent viscosities. The effects of surface tension on the inference decrease with increasing fluid viscosity (Fig.~\ref{fig:surf}B). As viscosity increases, initial conditions shift towards a Taylor drop regime with higher gravitational stresses and higher Bo (Fig.~\ref{fig:flowevol}A).

\subsection*{Effects of fluid opacity}
We query the effect of fluid opacity by adding $2.5$ mg/mL of blue dye to the viscosity standards and testing them in the CV viscometer. We collect a total of four videos for each viscosity. From the four videos, we initially use two, we augment, and we query the model for viscosity inference (Fig.~\ref{fig:color}). The opacity of the fluid has considerable effects on the processed frames (Fig. \ref{fig:color}B) and on the inferences (Fig.~\ref{fig:color}C - red), especially in the higher and lower viscosity ranges. In fact, below $\sim 1$ Pa.s and above $\sim 10$ Pa.s, the error on the inference becomes above 100\%. To reduce the effects of opacity, we take the extra two collected videos and add them to the training set. We retrain the model with the original 10 videos of the non-dyed 96 viscosities plus 2 videos of the dyed viscosity standards. We then query this retrained model with the two videos per viscosity of dyed fluid that the model has not been trained with. The inference with the retrained model is considerably improved (Fig.~\ref{fig:color}C - blue).  Errors are comparable to those of epistemic tests on clear fluids. Our results suggest that opacity plays a considerable role in inference and that the model must either be trained with fluids of different opacity or the image processing must be improved to reduce these effects.

\newpage
\begin{figure}
\centering
\includegraphics[width=17.8cm]{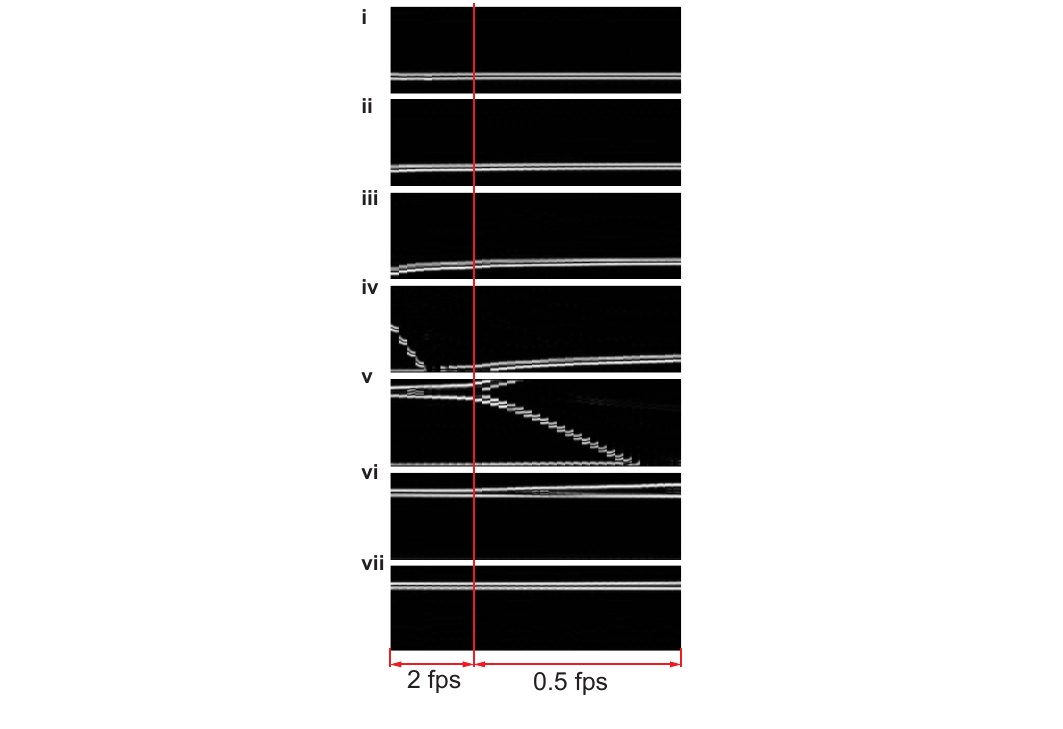}
\caption{Concatenated frames down-sampled to 2fps in the first 5 seconds and 0.5 fps for the remaining time. (i) 0.004 Pa.s, (ii) 0.04 Pa.s, (iii) 0.4 Pa.s, (iv) 3.9 Pa.s, (v) 34 Pa.s, (vi) 160 Pa.s, (vii) 1000 Pa.s.}
\label{fig:downsampled}
\end{figure}

\begin{figure}[!tbhp]
\centering
\includegraphics[width=7.42cm]{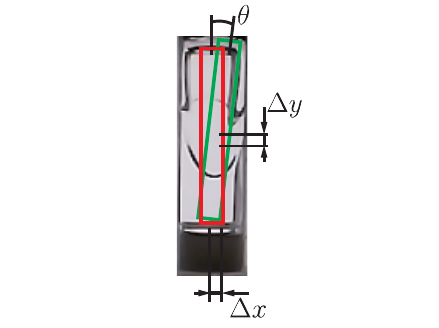}
\caption{Regions of interest picked for data augmentation. Ten additional regions of interest are picked from each video by randomly moving the original region of interest by $\Delta x$ and  $\Delta y$ between 0 and 3 pixels and randomly rotating the original region of interest by $\theta$ between 0 and 1 degree.}
\label{fig:augmentation}
\end{figure}

\begin{figure}[!tbhp]
\centering
\includegraphics[width=17.8cm]{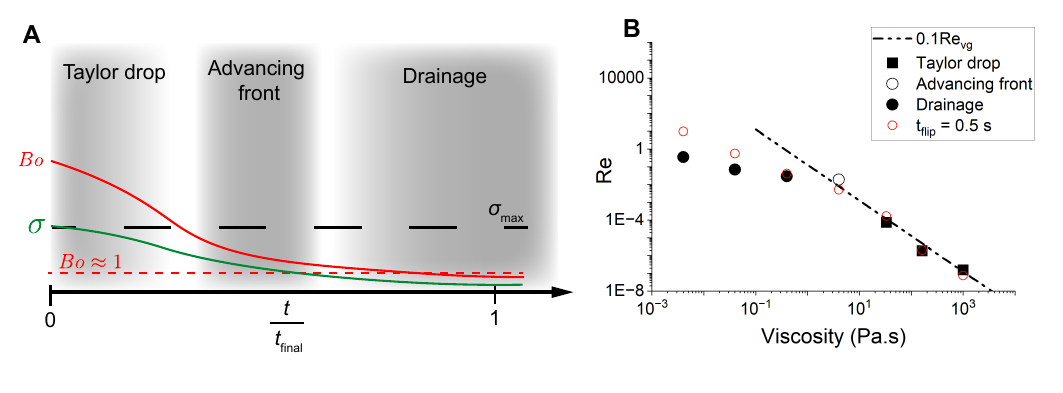}
\caption{
(A) Estimated evolution of the forcing stress, $\sigma$, and Bond number, Bo, during the inverted vial tests assuming an instantaneous flip. The trends shown are qualitative and not quantitative.
(B) Re$_\text{exp}$ as a function of viscosity for different $t_\text{flip}$. Black and white markers show Re$_\text{exp}$ for $t_\text{flip}=2$~s. The caption describes the regime observed when the vial reaches its fully-flipped condition, i.e., initial conditions. When the initial regime is a Taylor drop, i.e. high viscosities, Re$_\text{exp}$ approximates Re$_\text{VG}$ estimated from Buckingham $\Pi$ analysis (Eq.~\ref{eq:ReVG}). However, at the low viscosities, where initial conditions show drainage, Re$_\text{exp}$ deviates from Re$_\text{VG}$. Here, the measured velocities are considerably smaller than those predicted from Buckingham $\Pi$ analysis, resulting in the lower Re$_\text{exp}$. Shorter $t_\text{flip}$ does not affect high viscosities but results in considerably higher Re$_\text{exp}$ at the low viscosity regime. 
}
\label{fig:flowevol}
\end{figure}

\begin{figure}[!tbhp]
\centering
\includegraphics[width=17.8cm]{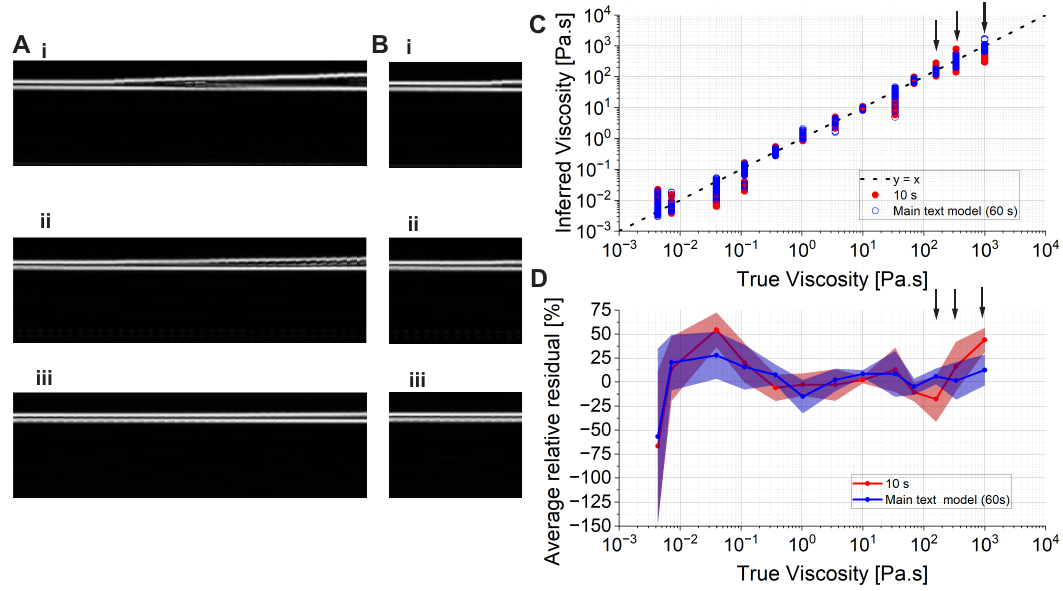}
\caption{Effects of video length on the epistemic accuracy. (A) Concatenated frames for 60 s and (B) 10 s video length for (i) 34 Pa.s, (ii) 160 Pa.s, (iii) 1000 Pa.s. (C) Inferred viscosities and (D) relative residuals. Shaded regions show standard deviations. Vertical arrows indicate the concatenated frames shown on the left.}
\label{fig:videolength}
\end{figure}

\begin{figure}[!tbhp]
\centering
\includegraphics[width=17.8cm]{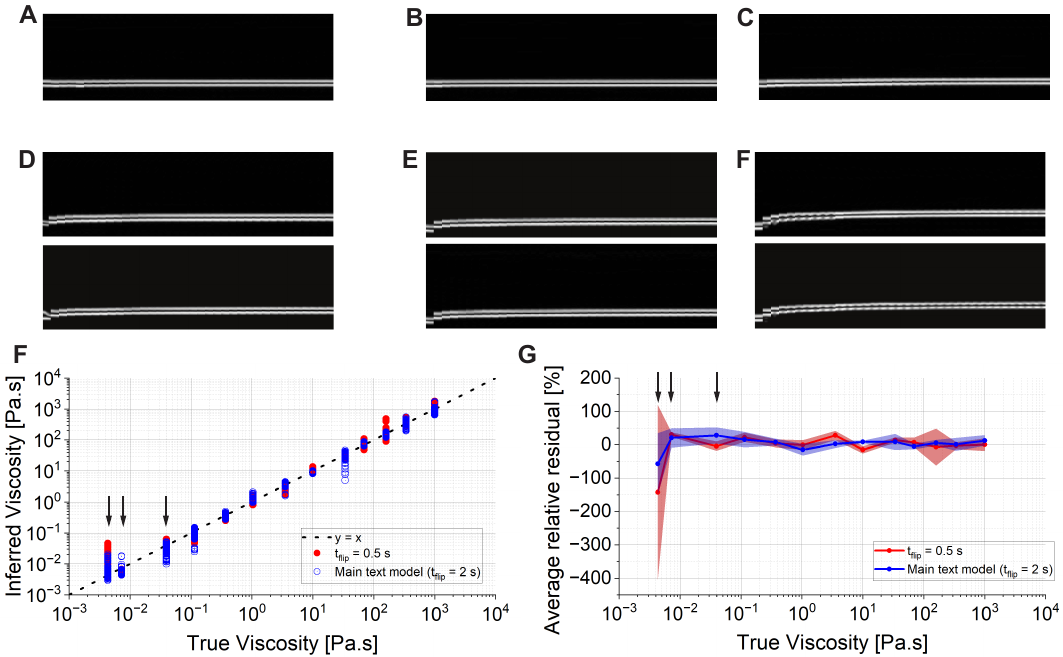}
\caption{Effects of flip speed on the epistemic accuracy. Concatenated frames for $t_\text{flip}=2$~s with viscosities (A) 0.004 Pa.s, (B) 0.04 Pa.s, (C) 0.4 Pa.s. Concatenated frames for two different flips and $t_\text{flip}=0.5$~s with viscosities (A) 0.004 Pa.s, (B) 0.04 Pa.s, (C) 0.4 Pa.s. (F) Inferred viscosities and (G) relative residuals. Shaded regions show standard deviations. Vertical arrows indicate the concatenated frames shown on the top.}
\label{fig:flipspeed}
\end{figure}

\begin{figure}[!tbhp]
\centering
\includegraphics[width=17.8cm]{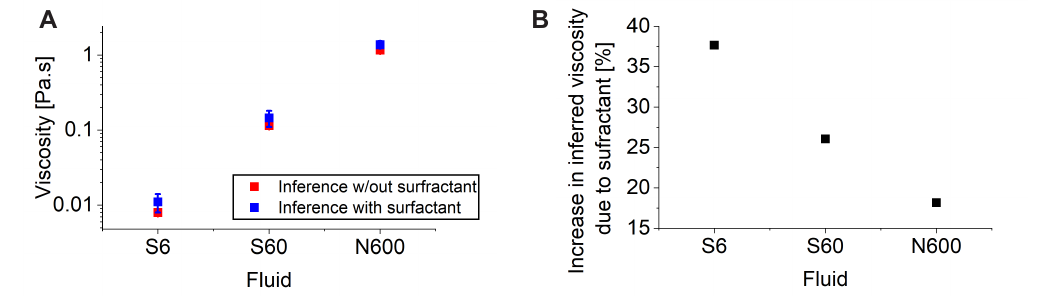}
\caption{Effect of surfactant on inferred viscosities. (A) Inferred viscosity from the neural network. (B) Increase in inferred viscosity from the addition of surfactant. The addition of surfactant results in an overestimation of the fluid viscosity. This effect is reduced as the viscosity of the fluid increases.}
\label{fig:surf}
\end{figure}

\begin{figure}[!tbhp]
\centering
\includegraphics[width=17.8cm]{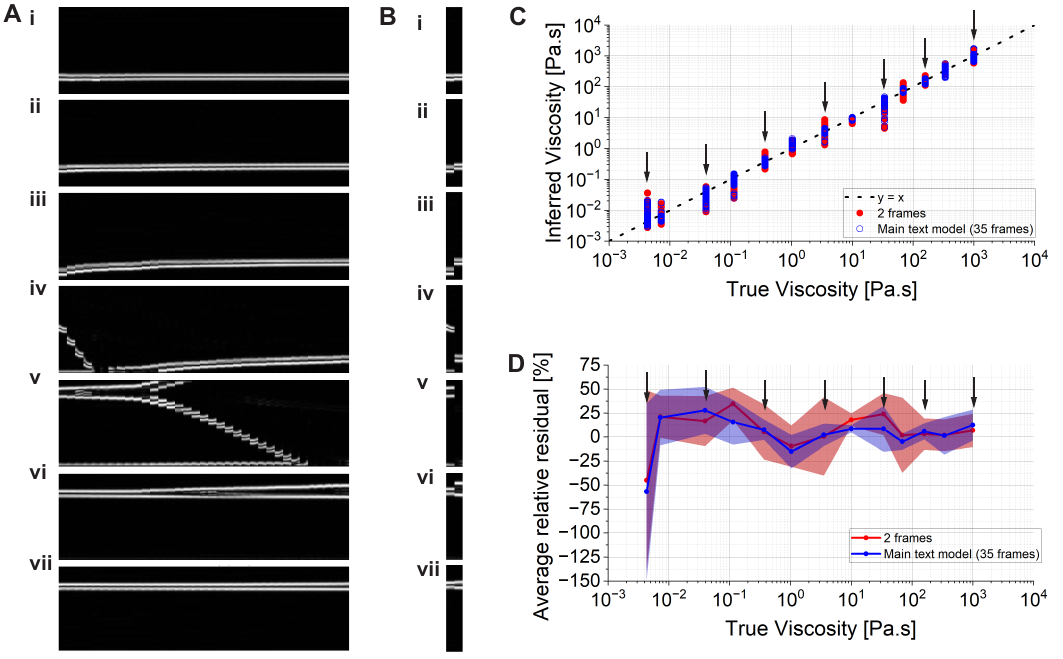}
\caption{Effects of frame rate on the epistemic accuracy. (A) Concatenated frames at 2~fps for first 5~s and 0.5~fps for remaining (35 frames total) and (B) 0.02~fps (2 frames total) for (i) 0.004 Pa.s, (ii) 0.04 Pa.s, (iii) 0.4 Pa.s, (iv) 3.9 Pa.s, (v) 34 Pa.s, (vi) 160 Pa.s, (vii) 1000 Pa.s. (C) Inferred viscosities and (D) relative residuals. Shaded regions show standard deviations. Vertical arrows indicate the concatenated frames shown on the left.}
\label{fig:framerate}
\end{figure}


\begin{figure}[!tbhp]
\centering
\includegraphics[width=17.8cm]{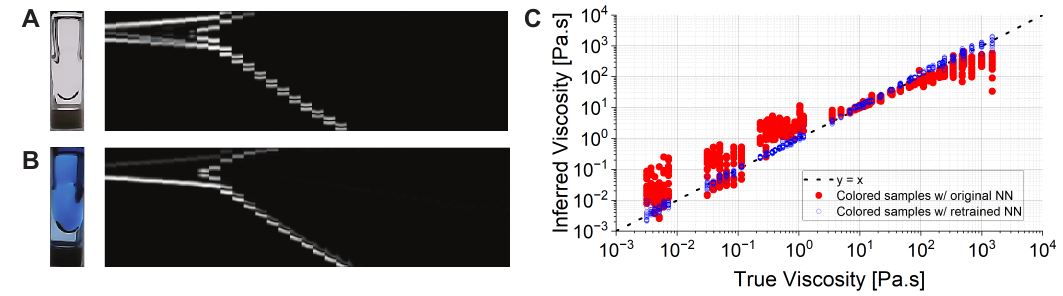}
\caption{Effect of fluid opacity on aleatoric accuracy. (A) Image of clear fluid after 30 s and corresponding concatenated frames. (B) Image of colored fluid after 30 s and corresponding concatenated frames. (C) Inferred viscosities.}
\label{fig:color}
\end{figure}

\begin{figure}[!tbhp]
\centering
\includegraphics[]{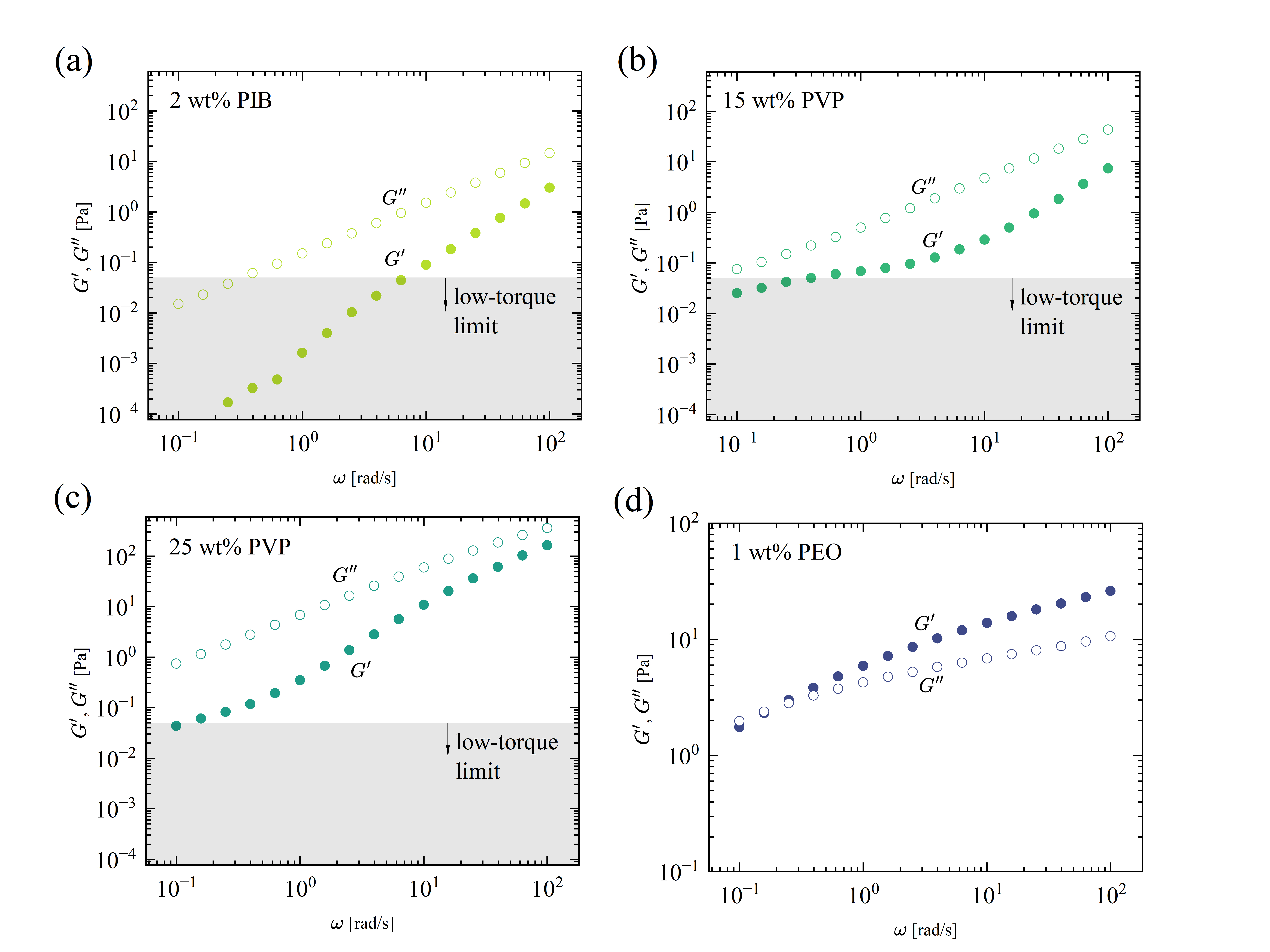}
\caption{Linear viscoelastic data of the non-Newtonian fluids (A) 2\%wt 500k polyisobutylene (PIB) in S6, (B) 15\%wt polyvinylpyrrolidone (PVP) in deionized water, (C) 25\%wt polyvinylpyrrolidone (PVP) in deionized water, (D) 1\%wt 8M polyethylene oxide (PEO) in deionized water obtained with 40mm parallel plate geometry (DHR-3, TA instruments, $h=1$ mm, $\gamma=0.1$ and $T=25^o$C). Data at low-frequency 2 wt\% PIB, 15 wt\% PVP and 25 wt\% PVP is unattainable due to the low-torque limit of the rheometer. The terminal region for these three fluids reveals a viscoelastic relaxation timescale by taking the ratio of $G'/(G''\omega)$ in the limit Deborah number De$\to 0$, which gives $\tau= 0.005, 0.007$ and  0.2 sec, respectively. For 1wt\% PEO solution, we don't observe a terminal region and the characteristic linear relaxation time can be approximated from $\tau\sim 1/\omega_c \sim4$ sec, where $\omega_c$ is the crossover frequency ($G'(\omega_c)\approx G''(\omega_c)$). The relaxation timescale $\tau$ provides a conservative upper bound of the retardation timescale $\tau_r$, as $\tau_r<\tau$~\cite{9}.
}
\label{fig:LVE}
\end{figure}

\begin{figure}[!tbhp]
\centering
\includegraphics[]{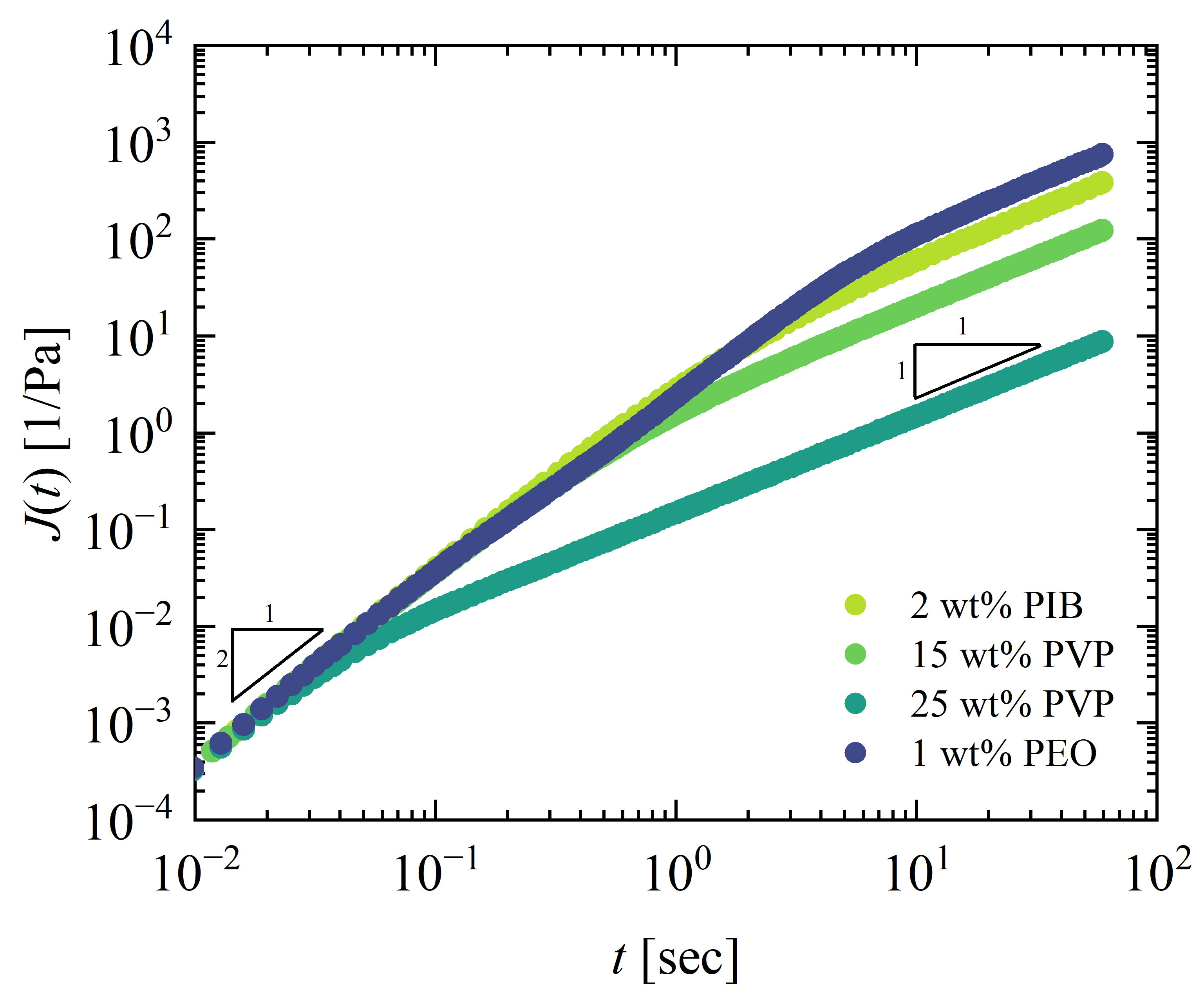}
\caption{Creep data for the non-Newtonian fluids were obtained using a 40~mm parallel plate geometry (DHR-3, TA Instruments) with a gap of \( h = 1 \)~mm at \( T = 25^\circ\text{C} \), and are plotted as apparent compliance \( J(t) \). For all fluids, we were unable to accurately determine the retardation timescale due to a short-time instrument acceleration artifact. This artifact, independent of the applied torque, appears as an apparent compliance scaling of \( J(t) \sim t^2 \), whereas the true viscoelastic material response would exhibit a slope between 0 and 1. For all tested fluids, the material response transitions to \( J(t) \sim t \) (viscous) behavior after the initial acceleration artifact. Thus, we can only estimate an upper bound for the retardation time, defined as the time when the creep response deviates from a slope of 2; this transition depends on the material viscosity. Based on this analysis, the retardation time for 2~wt\% PIB is less than 2~s, for 15~wt\% PVP is less than 0.9~s, for 25~wt\% PVP is less than about 0.1~s, and for 1~wt\% PEO is less than 5~s and consistent with the linear viscoelasticity data.
}
\label{fig:creep}
\end{figure}

\begin{figure}[!tbhp]
\centering
\includegraphics[width=18.8cm]{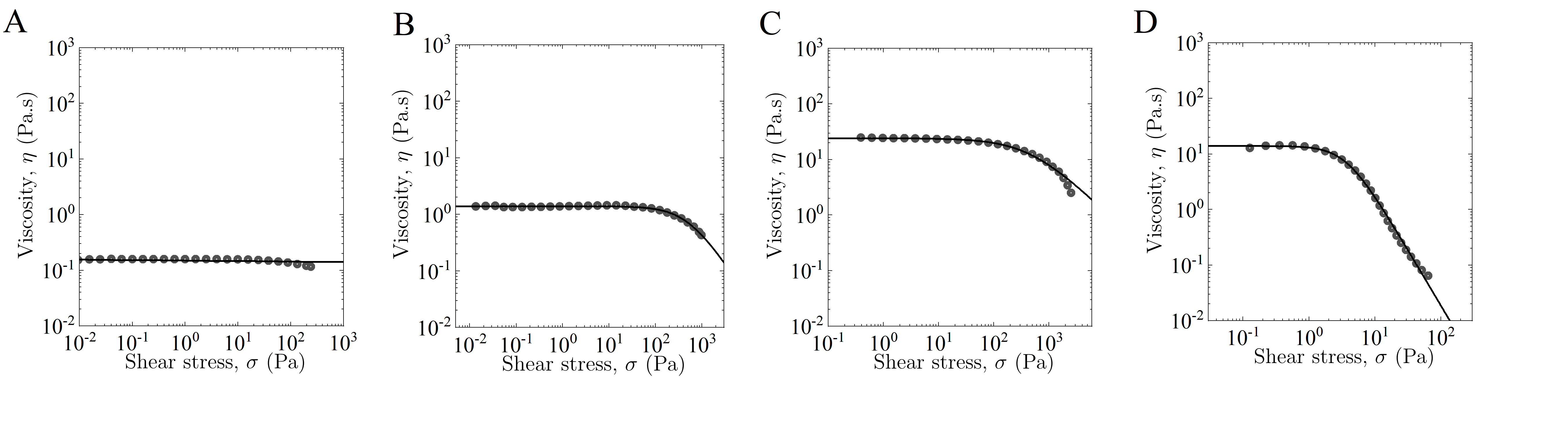}
\caption{Ellis model fitted to the steady shear data for (A) 2\%wt 500k polyisobutylene (PIB) in S6, (B) 15\%wt polyvinylpyrrolidone (PVP) in deionized water, (C) 25\%wt polyvinylpyrrolidone (PVP) in deionized water, (D) 1\%wt 8M polyethylene oxide (PEO) in deionized water.}
\label{fig:ellis}
\end{figure}

\begin{figure}[!tbhp]
\centering
\includegraphics[width=17.8cm]{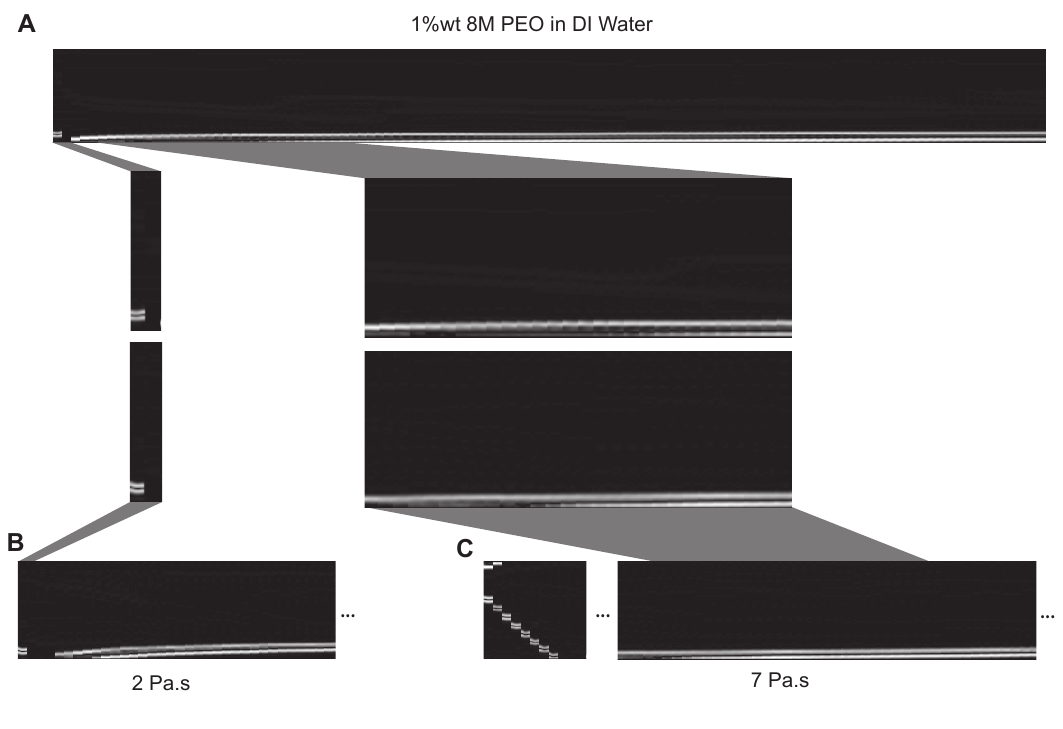}
\caption{(A) Concatenattion of first  frames for 1\%wt 8M PEO in DI water shown in Figure 6D. (B) Concatenated frames for Newtonian fluids with 2 Pa.s and (C) 7 Pa.s (not all frames are shown). The first two frames of 1\%wt 8M PEO in DI water resemble the first 2 frames of a 1~Pa·s Newtonian fluid. The subsequent 28 frames appear similar to later frames of a 7~Pa·s Newtonian fluid. }
\label{fig:nonNew}
\end{figure}

\begin{table}\centering
\caption{Viscosities and densities of training fluids}
\footnotesize
\begin{tabular}{|c|c|c|c|c|}
\hline
\textbf{Fluid}      & \textbf{Temperature} & \textbf{Viscosity {[}Pa.s{]}} & \textbf{Density {[}kg/m\textasciicircum{}3{]}} & \textbf{Density   measurement} \\ \hline
\textbf{S6}         & 25                   & 7.280E-03                     & 8.661E-01                                      & From spec. sheet               \\ \hline
\textbf{S6}         & 30                   & 6.055E-03                     & 8.627E-01                                      & Linear interpolation           \\ \hline
\textbf{S6}         & 35                   & 5.125E-03                     & 8.593E-01                                      & Linear interpolation           \\ \hline
\textbf{S6}         & 40                   & 4.304E-03                     & 8.559E-01                                      & From spec. sheet               \\ \hline
\textbf{S6}         & 45                   & 3.723E-03                     & 8.525E-01                                      & Linear interpolation           \\ \hline
\textbf{S6}         & 50                   & 3.188E-03                     & 8.491E-01                                      & From spec. sheet               \\ \hline
\textbf{S60}        & 25                   & 1.144E-01                     & 8.629E-01                                      & From spec. sheet               \\ \hline
\textbf{S60}        & 30                   & 8.510E-02                     & 8.598E-01                                      & Linear interpolation           \\ \hline
\textbf{S60}        & 35                   & 6.481E-02                     & 8.567E-01                                      & Linear interpolation           \\ \hline
\textbf{S60}        & 40                   & 5.008E-02                     & 8.536E-01                                      & From spec. sheet               \\ \hline
\textbf{S60}        & 45                   & 3.959E-02                     & 8.505E-01                                      & Interpolation                  \\ \hline
\textbf{S60}        & 50                   & 3.160E-02                     & 8.474E-01                                      & From spec. sheet               \\ \hline
\textbf{N600}       & 25                   & 1.187E+00                     & 8.437E-01                                      & From spec. sheet               \\ \hline
\textbf{N600}       & 30                   & 8.619E-01                     & 8.407E-01                                      & Linear interpolation           \\ \hline
\textbf{N600}       & 35                   & 6.395E-01                     & 8.378E-01                                      & Linear interpolation           \\ \hline
\textbf{N600}       & 40                   & 4.833E-01                     & 8.348E-01                                      & Linear interpolation           \\ \hline
\textbf{N600}       & 45                   & 3.709E-01                     & 8.318E-01                                      & Linear interpolation           \\ \hline
\textbf{N600}       & 50                   & 2.892E-01                     & 8.288E-01                                      & Linear interpolation           \\ \hline
\textbf{RTM24}      & 25                   & 1.038E+00                     & 8.727E-01                                      & DMA-501 density meter          \\ \hline
\textbf{RTM24}      & 30                   & 7.436E-01                     & 8.698E-01                                      & DMA-501 density meter          \\ \hline
\textbf{RTM24}      & 35                   & 5.471E-01                     & 8.668E-01                                      & DMA-501 density meter          \\ \hline
\textbf{RTM24}      & 40                   & 4.070E-01                     & 8.639E-01                                      & DMA-501 density meter          \\ \hline
\textbf{RTM24}      & 45                   & 3.082E-01                     & 8.610E-01                                      & Linear interpolation           \\ \hline
\textbf{RTM24}      & 50                   & 2.331E-01                     & 8.580E-01                                      & Linear interpolation           \\ \hline
\textbf{RT12500}    & 25                   & 1.330E+01                     & 9.702E-01                                      & Linear interpolation           \\ \hline
\textbf{RT12500}    & 30                   & 1.202E+01                     & 9.656E-01                                      & From spec. sheet               \\ \hline
\textbf{RT12500}    & 35                   & 1.092E+01                     & 9.612E-01                                      & Linear interpolation           \\ \hline
\textbf{RT12500}    & 40                   & 9.955E+00                     & 9.568E-01                                      & From spec. sheet               \\ \hline
\textbf{RT12500}    & 45                   & 9.108E+00                     & 9.524E-01                                      & Linear interpolation           \\ \hline
\textbf{RT12500}    & 50                   & 8.348E+00                     & 9.480E-01                                      & Linear interpolation           \\ \hline
\textbf{N8000}      & 25                   & 2.151E+01                     & 8.856E-01                                      & DMA-501 density meter          \\ \hline
\textbf{N8000}      & 30                   & 1.450E+01                     & 8.829E-01                                      & DMA-501 density meter          \\ \hline
\textbf{N8000}      & 35                   & 9.949E+00                     & 8.801E-01                                      & DMA-501 density meter          \\ \hline
\textbf{N8000}      & 40                   & 6.929E+00                     & 8.774E-01                                      & DMA-501 density meter          \\ \hline
\textbf{N8000}      & 45                   & 4.907E+00                     & 8.747E-01                                      & Linear interpolation           \\ \hline
\textbf{N8000}      & 50                   & 3.540E+00                     & 8.719E-01                                      & Linear interpolation           \\ \hline
\textbf{N30000}     & 25                   & 6.894E+01                     & 8.949E-01                                      & From spec. sheet               \\ \hline
\textbf{N30000}     & 30                   & 4.655E+01                     & 8.922E-01                                      & Linear interpolation           \\ \hline
\textbf{N30000}     & 35                   & 3.200E+01                     & 8.896E-01                                      & Linear interpolation           \\ \hline
\textbf{N30000}     & 40                   & 2.230E+01                     & 8.869E-01                                      & Linear interpolation           \\ \hline
\textbf{N30000}     & 45                   & 1.576E+01                     & 8.842E-01                                      & Linear interpolation           \\ \hline
\textbf{N30000}     & 50                   & 1.135E+01                     & 8.815E-01                                      & Linear interpolation           \\ \hline
\textbf{N62000}     & 25                   & 2.038E+02                     & 8.993E-01                                      & DMA-501 density meter          \\ \hline
\textbf{N62000}     & 30                   & 1.373E+02                     & 8.966E-01                                      & DMA-501 density meter          \\ \hline
\textbf{N62000}     & 35                   & 9.417E+01                     & 8.941E-01                                      & DMA-501 density meter          \\ \hline
\textbf{N62000}     & 40                   & 6.440E+01                     & 8.915E-01                                      & DMA-501 density meter          \\ \hline
\textbf{N62000}     & 45                   & 4.547E+01                     & 8.889E-01                                      & Linear interpolation           \\ \hline
\textbf{N62000}     & 50                   & 3.396E+01                     & 8.863E-01                                      & Linear interpolation           \\ \hline

\end{tabular}
\label{Table:Table}
\end{table}

\begin{table}\centering
\caption*{Tables S1: Viscosities and densities of training fluids (continued)}
\footnotesize
\begin{tabular}{|c|c|c|c|c|}
\hline
\textbf{Fluid}      & \textbf{Temperature} & \textbf{Viscosity {[}Pa.s{]}} & \textbf{Density {[}kg/m\textasciicircum{}3{]}} & \textbf{Density   measurement} \\ \hline
\textbf{N190000}    & 25                   & 5.080E+02                     & 8.993E-01                                      & From N62000                    \\ \hline
\textbf{N190000}    & 30                   & 3.382E+02                     & 8.966E-01                                      & From N62000                    \\ \hline
\textbf{N190000}    & 35                   & 2.284E+02                     & 8.941E-01                                      & From N62000                    \\ \hline
\textbf{N190000}    & 40                   & 1.590E+02                     & 8.915E-01                                      & From N62000                    \\ \hline
\textbf{N190000}    & 45                   & 1.113E+02                     & 8.889E-01                                      & From N62000                    \\ \hline
\textbf{N190000}    & 50                   & 7.942E+01                     & 8.863E-01                                      & From N62000                    \\ \hline
\textbf{N450000}    & 25                   & 1.484E+03                     & 8.993E-01                                      & From N62000                    \\ \hline
\textbf{N450000}    & 30                   & 1.002E+03                     & 8.966E-01                                      & From N62000                    \\ \hline
\textbf{N450000}    & 35                   & 6.894E+02                     & 8.941E-01                                      & From N62000                    \\ \hline
\textbf{N450000}    & 40                   & 4.827E+02                     & 8.915E-01                                      & From N62000                    \\ \hline
\textbf{N450000}    & 45                   & 3.435E+02                     & 8.889E-01                                      & From N62000                    \\ \hline
\textbf{N450000}    & 50                   & 2.465E+02                     & 8.863E-01                                      & From N62000                    \\ \hline
\textbf{RTM30}      & 25                   & 5.007E+00                     & 8.817E-01                                      & DMA-501 density meter          \\ \hline
\textbf{RTM30}      & 30                   & 3.480E+00                     & 8.789E-01                                      & DMA-501 density meter          \\ \hline
\textbf{RTM30}      & 35                   & 2.468E+00                     & 8.761E-01                                      & DMA-501 density meter          \\ \hline
\textbf{RTM30}      & 40                   & 1.776E+00                     & 8.732E-01                                      & DMA-501 density meter          \\ \hline
\textbf{RTM30}      & 45                   & 1.310E+00                     & 8.704E-01                                      & Linear interpolation           \\ \hline
\textbf{RTM30}      & 50                   & 9.851E-01                     & 8.675E-01                                      & Linear interpolation           \\ \hline
\textbf{RTM32}      & 25                   & 6.868E+00                     & 8.819E-01                                      & DMA-501 density meter          \\ \hline
\textbf{RTM32}      & 30                   & 4.911E+00                     & 8.791E-01                                      & DMA-501 density meter          \\ \hline
\textbf{RTM32}      & 35                   & 3.514E+00                     & 8.762E-01                                      & DMA-501 density meter          \\ \hline
\textbf{RTM32}      & 40                   & 2.572E+00                     & 8.735E-01                                      & DMA-501 density meter          \\ \hline
\textbf{RTM32}      & 45                   & 1.903E+00                     & 8.707E-01                                      & Linear interpolation           \\ \hline
\textbf{RTM32}      & 50                   & 1.430E+00                     & 8.679E-01                                      & Linear interpolation           \\ \hline
\textbf{Honey}      & 25                   & 9.441E+00                     & 1.419E+00                                      & DMA-501 density meter          \\ \hline
\textbf{Honey}      & 30                   & 5.938E+00                     & 1.416E+00                                      & DMA-501 density meter          \\ \hline
\textbf{Honey}      & 35                   & 3.977E+00                     & 1.413E+00                                      & DMA-501 density meter          \\ \hline
\textbf{Honey}      & 40                   & 2.735E+00                     & 1.409E+00                                      & DMA-501 density meter          \\ \hline
\textbf{Honey}      & 45                   & 1.999E+00                     & 1.406E+00                                      & Linear interpolation           \\ \hline
\textbf{Honey}      & 50                   & 1.412E+00                     & 1.402E+00                                      & Linear interpolation           \\ \hline
\textbf{Glycerol}     & 25                   & 8.120E-01                     & 1.260E+00                                      & DMA-501 density meter          \\ \hline
\textbf{Glycerol}     & 30                   & 5.270E-01                     & 1.258E+00                                      & DMA-501 density meter          \\ \hline
\textbf{Glycerol}     & 35                   & 3.571E-01                     & 1.254E+00                                      & DMA-501 density meter          \\ \hline
\textbf{Glycerol}     & 40                   & 2.465E-01                     & 1.252E+00                                      & DMA-501 density meter          \\ \hline
\textbf{Glycerol}     & 45                   & 1.762E-01                     & 1.249E+00                                      & Linear interpolation           \\ \hline
\textbf{Glycerol}     & 50                   & 1.270E-01                     & 1.246E+00                                      & Linear interpolation           \\ \hline
\textbf{CornSyrup1} & 25                   & 7.179E+00                     & 1.391E+00                                      & DMA-501 density meter          \\ \hline
\textbf{Corn Syrup 1} & 30                   & 4.600E+00                     & 1.386E+00                                      & DMA-501 density meter          \\ \hline
\textbf{Corn Syrup 1} & 35                   & 3.086E+00                     & 1.382E+00                                      & DMA-501 density meter          \\ \hline
\textbf{Corn Syrup 1} & 40                   & 2.129E+00                     & 1.378E+00                                      & DMA-501 density meter          \\ \hline
\textbf{Corn Syrup 1} & 45                   & 1.482E+00                     & 1.374E+00                                      & Linear interpolation           \\ \hline
\textbf{Corn Syrup 1} & 50                   & 1.059E+00                     & 1.370E+00                                      & Linear interpolation           \\ \hline
\textbf{Corn Syrup 2} & 25                   & 7.473E+01                     & 1.391E+00                                      & From Corn Syrup 1               \\ \hline
\textbf{Corn Syrup 2} & 30                   & 4.149E+01                     & 1.386E+00                                      & From Corn Syrup 1               \\ \hline
\textbf{Corn Syrup 2} & 35                   & 2.483E+01                     & 1.382E+00                                      & From Corn Syrup 1               \\ \hline
\textbf{Corn Syrup 2 } & 40                   & 1.591E+01                     & 1.378E+00                                      & From Corn Syrup 1               \\ \hline
\textbf{Corn Syrup 2} & 45                   & 1.080E+01                     & 1.374E+00                                      & From Corn Syrup 1               \\ \hline
\textbf{Corn Syrup 2} & 50                   & 7.776E+00                     & 1.370E+00                                      & From Corn Syrup 1               \\ \hline

\end{tabular}
\end{table}

\normalsize
\begin{table}\centering
\caption{Model fitting parameters for the Ellis model}
\begin{tabular}{|c|c|c|c|c|}
\hline
Fluid name & $\eta_0$ [Pa.s] & $k$ [1/Pa] & a & $\sigma_{crit}=1/k$ [Pa]\\ \hline

2 wt\% 500k PIB in S6 & 0.24 & 1.7E-09 & 1.02 & 5.8E08 \\ \hline

15 wt\% PVP in DI water & 1.4 & 0.0019 & 2.25 & 526.3 \\ \hline

25 wt\% PVP in DI water & 24.1 & 0.0021 & 1.98 & 476.2 \\ \hline

1 wt\% 8M PEO in DI water & 13.9 & 0.27 & 3 & 3.7 \\ \hline
\end{tabular}
\label{Table:non-Newtonain parameters}
\end{table}




\FloatBarrier  